\documentclass{svproc}
\usepackage{url}

\usepackage{xcolor}
\usepackage{subcaption}
\usepackage{graphicx}

\usepackage[utf8]{inputenc}

\begin{document}
\mainmatter              % start of a contribution

\title{Can WhatsApp Counter Misinformation by Limiting Message Forwarding?}

%\title{Can WhatsApp Counter Misinformation Campaigns by Limiting its Virality Features?}

%
%\titlerunning{Understanding the Limits of Information Dissemination on Whatsapp}  % abbreviated title (for running head)
\author{Philipe de Freitas Melo\inst{1} \and Carolina Coimbra Vieira\inst{1} \and Kiran Garimella\inst{2} \and \\ Pedro O. S. Vaz de Melo\inst{1} \and Fabrício Benevenuto\inst{1}}
\authorrunning{Philipe Melo et al.} % abbreviated author list (for running head)
\tocauthor{Philipe de Freitas Melo, Carolina Coimbra Vieira, Kiran Garimella, Pedro O. S. Vaz de Melo, and  Fabrício Benevenuto}
\institute{Departamento de Ciência da Computação - UFMG, Belo Horizonte MG, Brazil,\\
\email{{philipe, carolcoimbra, olmo, fabricio}@dcc.ufmg.br}
\and
Institute for Data, Society and Systems - MIT, USA, \\
\email{{garimell}@mit.edu}}

\maketitle              % typeset the title of the contribution

\begin{abstract}

WhatsApp is the most popular messaging app in the world.
The closed nature of the app, in addition to the ease of transferring multimedia and sharing information to large-scale groups make WhatsApp unique among other platforms, where an anonymous encrypted messages can become viral, reaching multiple users in a short period of time. 
The personal feeling and immediacy of messages directly delivered to the user's phone on WhatsApp was extensively abused to spread unfounded rumors and create misinformation campaigns during recent elections in Brazil and India. 
WhatsApp has been deploying measures to mitigate this problem, such as reducing the limit for forwarding a message to at most five users at once. 
Despite the welcomed effort to counter the problem, there is no evidence so far on the real effectiveness of such restrictions. In this work, we propose a methodology to evaluate the effectiveness of such measures on the spreading of misinformation circulating on WhatsApp. We use an epidemiological model and real data gathered from WhatsApp in Brazil, India and Indonesia to assess the impact of limiting virality features in this kind of network. Our results suggest that the current efforts deployed by WhatsApp can offer significant delays on the information spread, but they are ineffective in blocking the propagation of misinformation campaigns through public groups when the content has a high viral nature.

%Our results suggest that the current efforts deployed by WhatsApp can offer delays on the information spread, but are ineffective in blocking the propagation of misinformation campaigns in public groups. 

\keywords{WhatsApp, Misinformation, Fake news, Virality, Epidemics, Diffusion, Susceptible-Exposed-Infected,  Complex Network}
\end{abstract}

\section{Introduction}
\label{sec:intro}
\vspace{-0.2cm}

Messaging applications, such as WhatsApp, Facebook messenger, Telegram and Viber have gained a significant role in the daily lives of smartphone users.
WhatsApp is the most popular app, with over 1 billion active users\footnote{\url{https://blog.whatsapp.com/10000631/Connecting anuser-users-all-days}}. Besides being widely used to keep in touch with friends \& family, run businesses, read news \& get informed, WhatsApp has become an important platform for information dissemination and social mobilization, especially in Brazil, India and Southeast Asia~\cite{resendewww19}. 

There are a few key features that make WhatsApp unique among other platforms. 
First, WhatsApp allows the connection among like-minded individuals through chat groups. These chat groups have a limit of 256 users and can be private or public.
In the case of private groups, new members must be added by a member who assumes the role of group administrator. For public groups, the access is by invitation links that could be shared to anyone or be available on the Web. These public groups often come up to discuss hobbies and passions, but also specific topics such as health, education, and politics. Although the majority of groups are private, set up among people who share a social relationship (e.g., family, friends, workmates) public groups have been a catalyzing feature for the purpose of information diffusion: most of their members are strangers to each other.
This is evident in countries like Brazil, where a survey reported that 76\% of WhatsApp users are part of groups, 58\% participate in groups with people they do not know, and 18\% of these groups discuss politics~\cite{reuters2019report}. 
For this reason, public groups can act as a shortcut for information to directly traverse distant parts of the underlying social network structure via a clique of weak ties, broadening and accelerating information dissemination~\cite{Bakshy:2012:RSN}.

Furthermore, the app has two sharing functions: broadcast, in which a contact list can be created to send messages to up to 256 contacts (users or groups) at once and forward, that a single message received can be forwarded to other 5 contacts (users or groups). Those characteristics allow the message to travel long distances by the network, but the end-to-end encryption makes it difficult to identify the source and track the spread of the messages.
Because of these peculiarities, WhatsApp generated a controversy related to its anonymity and virality characteristics. This conflict is due the fact that we can view WhatsApp in two different ways, such as a technology company, or as a media platform. As a technology platform, it ensures user anonymity and security by encrypting your data. As a media platform, it transmits information and disseminates content in large-scale. Thus, messages sent anonymously reach thousands of people quickly and without any ethical or legal regulation of this disseminated content, promoting, for example, disinformation campaigns. The massive spread of misinformation and rumors~\cite{arun2019whatsapp} led to requests from both the national governments\footnote{\url{https://www.latimes.com/world/la-fg-india-whatsapp-2019-story.html}} 
%and later the Brazillian\footnote{\url{www.nytimes.com/2018/10/17/opinion/brazil-election-fake-news-whatsapp.html}} 
towards altering features that allow the platform to be abused to spread misinformation at scale.
This resulted in WhatsApp implementing restrictions on the way messages are forwarded\footnote{\url{blog.whatsapp.com/10000647/More-changes-to-forwarding}} by reducing the limit for forwarding content to at most 5 users/groups at a time. 
However, there are no studies that investigate the impact of these limitations or whether the numbers chosen are sufficient to deal with the spread of viral content.

In this work, we evaluate the dynamics of the spread of (mis)information on a network of public WhatsApp groups. 
We focus on the mass communication features of public chat groups and the forwarding/broadcasting of messages. 
More specifically, we study the anatomy of this emerging social network and comprehend its peculiarities to answer the question of how the forwarding tools contribute to the virality of (mis)information and whether system limitations are capable of preventing the spread of content. 
We also propose some hints on how the problem of large-scale dissemination can be countered.

The rest of the paper is organized as follows. In Section~\ref{sec:related} we describe the related work. In Section~\ref{sec:datasets} we describe the WhatsApp data used in this paper together with the methodology used to collect it. An initial characterization of the data is shown in Section~\ref{sec:spreading}. In Section~\ref{sec:networks}, we reconstruct a network from the collected data and we compare its characteristics with other real and synthetic networks. In Section~\ref{sec:sei}, we execute several experiments to measure the virality of a potential misinformation within these networks via the Susceptible-Exposed-Infected (SEI) epidemiological model~\cite{guihua2004global}. 
Finally, in Section~\ref{sec:conclusions}, we discuss our findings and final conclusions from the analysis.

\section{Related Work}
\label{sec:related}
\vspace{-0.1cm}

Recently, there have been numerous research studies reporting misinformation campaigns on social networks~\cite{bessi2016social,lazer2018science}. 
This includes popular platforms like Facebook, where, Ribiero \textit{et al.}~\cite{ribeiroFAT2019} evaluated the use of the Facebook advertising platform to carry out political campaigns that exploit targeted marketing as a means of disseminating false advertisements or on divisible themes. 
There are also reports of attempts to manipulate political discourse with the use of social bots and even state-sponsored trolls~\cite{bessi2016social,ferrara2017disinformation,zannettou2019disinformation}.  

However, only recently, social message applications, such as WhatsApp were reportedly a means of abuse by misinformation campaigns~\cite{resendewww19,resendewebsci19,philipe2019whatsapp,bursztyn2019thousands}. Particularly, Resende \textit{et al.}~\cite{resendewww19} analyzed the dissemination of different kinds of content on WhatsApp, such as images, audio and videos, finding a large amount of misinformation in the form of memes and fake images. 
Resende \textit{et al.}~\cite{resendewebsci19} provide an in-depth characterization of textual messages, showing that misinformation tends to be more viral, i.e., these messages are shared more times, by a larger number of users, and in more public groups.
Bursztyn \textit{et al.} showed that right-wing WhatsApp groups in Brazil were  more active and engaged in spreading political content in WhatsApp along the 2018 Brazilian elections, in comparison with left-wing groups. Melo \textit{et al.}~\cite{philipe2019whatsapp} developed a system to help fact checkers, providing them a sample of the most popular  images, messages, URLs, audios and videos shared hundreds of public groups in Brazil and India. 
This system has been used by Comprova, a collaborative journalism project from First Draft focused on verifying questionable stories published on social media and WhatsApp during the 12 weeks leading up to the Brazilian 2018 presidential election~\cite{firstdraf2019report}.
Our work is complementary to the above efforts as we investigate how limitations on virality features such as limits on message forwarding recently deployed by WhatsApp, are effective in mitigating misinformation campaigns. 

\section{Datasets}
\label{sec:datasets}
\vspace{-0.1cm}

Since chat groups on WhatsApp are mostly private, they are much harder to monitor than Facebook or Twitter discussions. Because of that, we use recent tools developed by Garimella and Tyson~\cite{garimella2018whatsapp} to get access to messages posted on WhatsApp public groups. Given a set of invitation links to public groups, we automatically join these groups and save all data coming from them. %For this work, 
We selected groups from Brazil, India and Indonesia dedicated to political discussions. These groups have a large flow of content and are mostly operated by individuals affiliated with political parties, or local community leaders. We monitored the groups during the electoral campaign period and, for each message, we extracted the following information:
(i) the country where the message was posted, (ii) name of the group the message was posted, (iii) user ID, (iv) timestamp and, when available, (v) the attached multimedia files (e.g. images, audio and videos).

As images usually flow unaltered across the network, they are easier to track than text messages. Thus, we choose to use the images posted on WhatsApp to analyse and understand how a single piece of content flows across the network. To calculate a fingerprint for every image, we follow the same strategy of \cite{resendewww19}, using the Perceptual Hashing (pHash) algorithm
to group together sets of images with the same content. Since similar images have the same hash value, we can count its popularity and track its spreading across the network. In total for all three countries, 784k unique image objects were tracked.

For all three countries, we analyzed the data around the election day, 60 days before and 15 after. We kept the same time span for the three countries to ease the comparison among them. The dataset overview and the total number of distinct images are described in Table \ref{tab:dataset}. As expected, Brazil and India have a much larger volume of data %(258k unique images from Brazil and 509k in India)
shared on WhatsApp compared to Indonesia, %(16k unique images), 
as they have much more groups and users registered in our data collection system.

\textbf{Data Limitations: }
Our methodology gathers a large dataset from public groups, but it is known that most of WhatsApp conversations occurs in private channels. A key limitation of our work is that our results reflect only users and content that circulate on the public layer of WhatsApp. We note, however, that there is evidence that suggests that public groups make up the key backbone of the misinformation campaigns on WhatsApp.\footnote{\url{https://www.bbc.com/news/world-asia-india-47797151}}
First, they are focused on political activism, where most of the shared content contain misinformation. For example, a fact checking agency in Brazil checked the top 61 images shared in these groups, finding that only 10\% of them are true~\cite{resendewww19}. There is also evidence of the use of automatic tools to flood WhatsApp public groups with political content\footnote{https://www.bbc.com/news/technology-45956557}. Then, the users in those groups would be responsible to amplify the misinformation campaign and propagate it to the private part of the network.\footnote{\url{https://time.com/5512032/whatsapp-india-election-2019/}} 
 
Nevertheless, this project brings a considerable amount of data 
that can help to elucidate how WhatsApp is being abused for mass communication
and the amplification backbone composed by public groups that distribute messages in bulk for thousands of users.
At the least, our results provide a `lower bound' on the ability of messages to spread on WhatsApp, since the network we consider is a subset of the entire WhatsApp network.

\begin{table}[t]
\centering
\caption{Overview of the datasets.}
%\vspace{-0.2cm}
\begin{tabular}{l|l|l|l|l|l|}
\cline{2-6}
\multicolumn{1}{c|}{}           & \multicolumn{1}{c|}{\#Users} & \multicolumn{1}{c|}{\#Groups} & \multicolumn{1}{c|}{\begin{tabular}[c]{@{}c@{}}Unique\\ Images\end{tabular}} & \multicolumn{1}{c|}{\begin{tabular}[c]{@{}c@{}}Total \\ Images\end{tabular}} & \multicolumn{1}{c|}{\begin{tabular}[c]{@{}c@{}}Period\\ $\sim$2,5 months\end{tabular}} \\ \hline
\multicolumn{1}{|l|}{Brazil}    & 17,465                       & 414                           & 258k                                                                         & 416k                                                                         & 2018/08/15 - 2018/11/01                                                               \\ \hline
\multicolumn{1}{|l|}{India}     & 362,739                      & 5,839                         & 509k                                                                         & 810k                                                                         & 2019/03/15 - 2019/06/01                                                                \\ \hline
\multicolumn{1}{|l|}{Indonesia} & 8,388                        & 217                           & 16k                                                                          & 21k                                                                          & 15/03/2019 - 2019/06/01                                                                \\ \hline
\end{tabular}
\label{tab:dataset}
\vspace{-0.2cm}
\end{table}

\section{Spreading Coverage and Dynamics}
\label{sec:spreading}
\vspace{-0.1cm}

Since we are able to track all occurrence of a given image, we can see the coverage and dynamics of spreading of these images in our data. To evaluate spreading metrics regarding time and coverage, we only consider the images that were posted in at least two groups, since we cannot see the effect of spreading of images only shared in a single group. This set consists of 2,384 images in Indonesia, 103,031 images in Brazil and 44,731 images for India, which represents approximately 20\% of the images for each country.

\begin{figure}[t]
	\centering%\hspace{-0.3cm}
    \begin{subfigure}[b]{0.24\textwidth}
        \includegraphics[width=\textwidth]{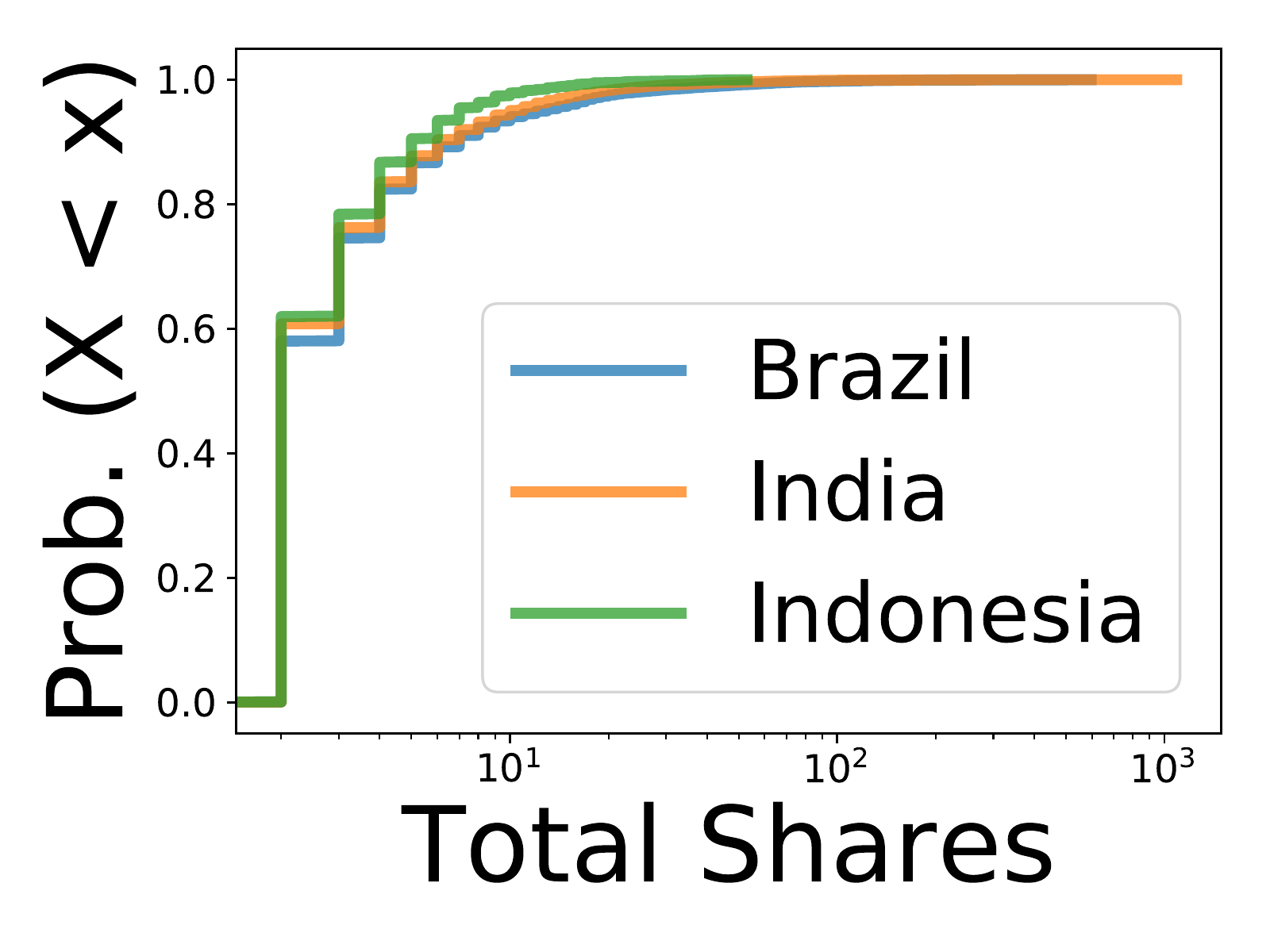}
    	\caption{Total shares.}
    	\label{fig:totalshares}
	%\qquad
	\end{subfigure}%\hspace{-0.2cm}
	\begin{subfigure}[b]{0.24\textwidth}
	    \includegraphics[width=\textwidth]{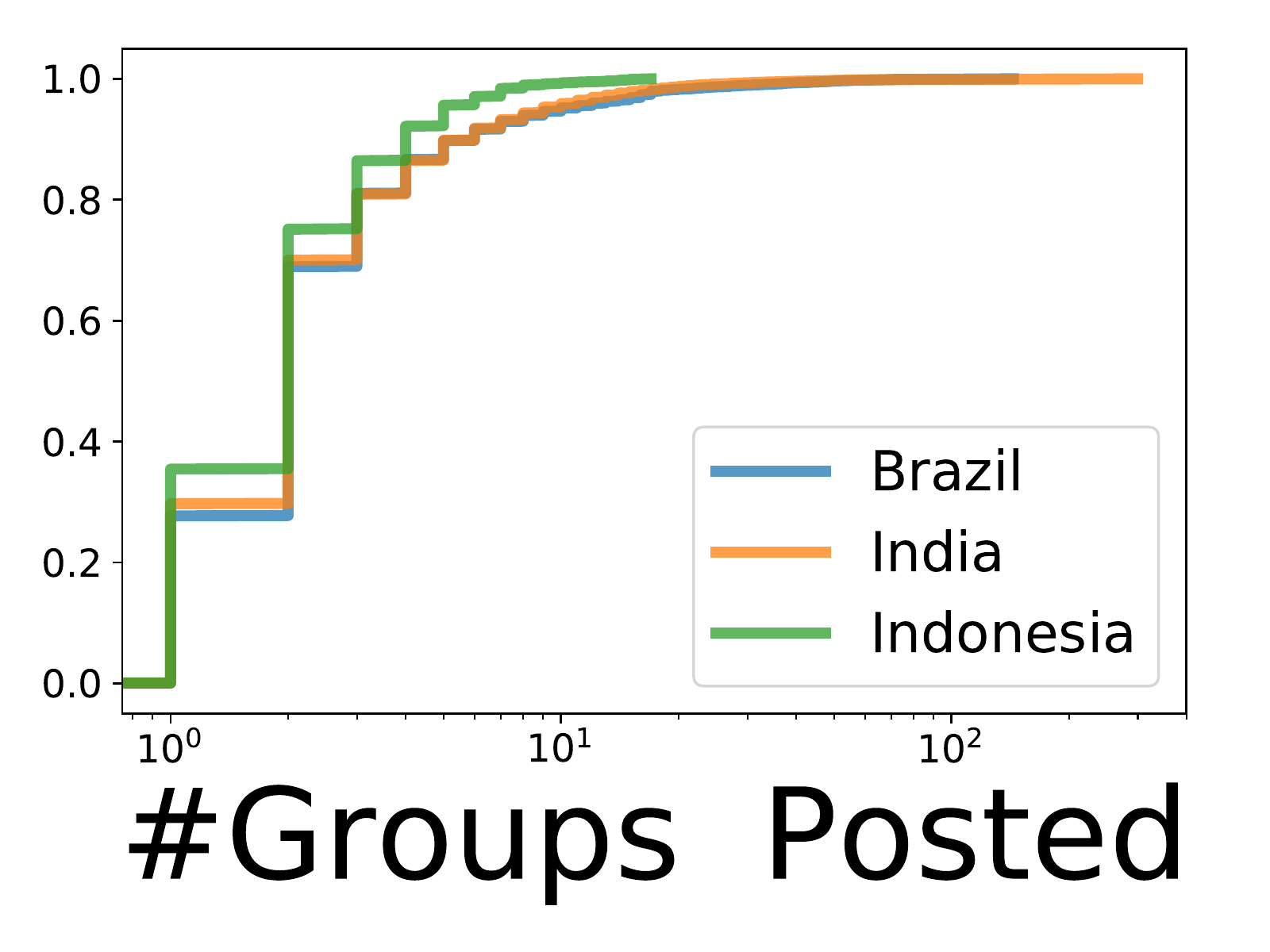}
	    \caption{Total groups}
	    \label{fig:totalgroups}
	%\qquad
	\end{subfigure}%\hspace{-0.2cm}
	%\vspace{-0.2cm}
	\begin{subfigure}[b]{0.24\textwidth}
		\includegraphics[width=\textwidth]{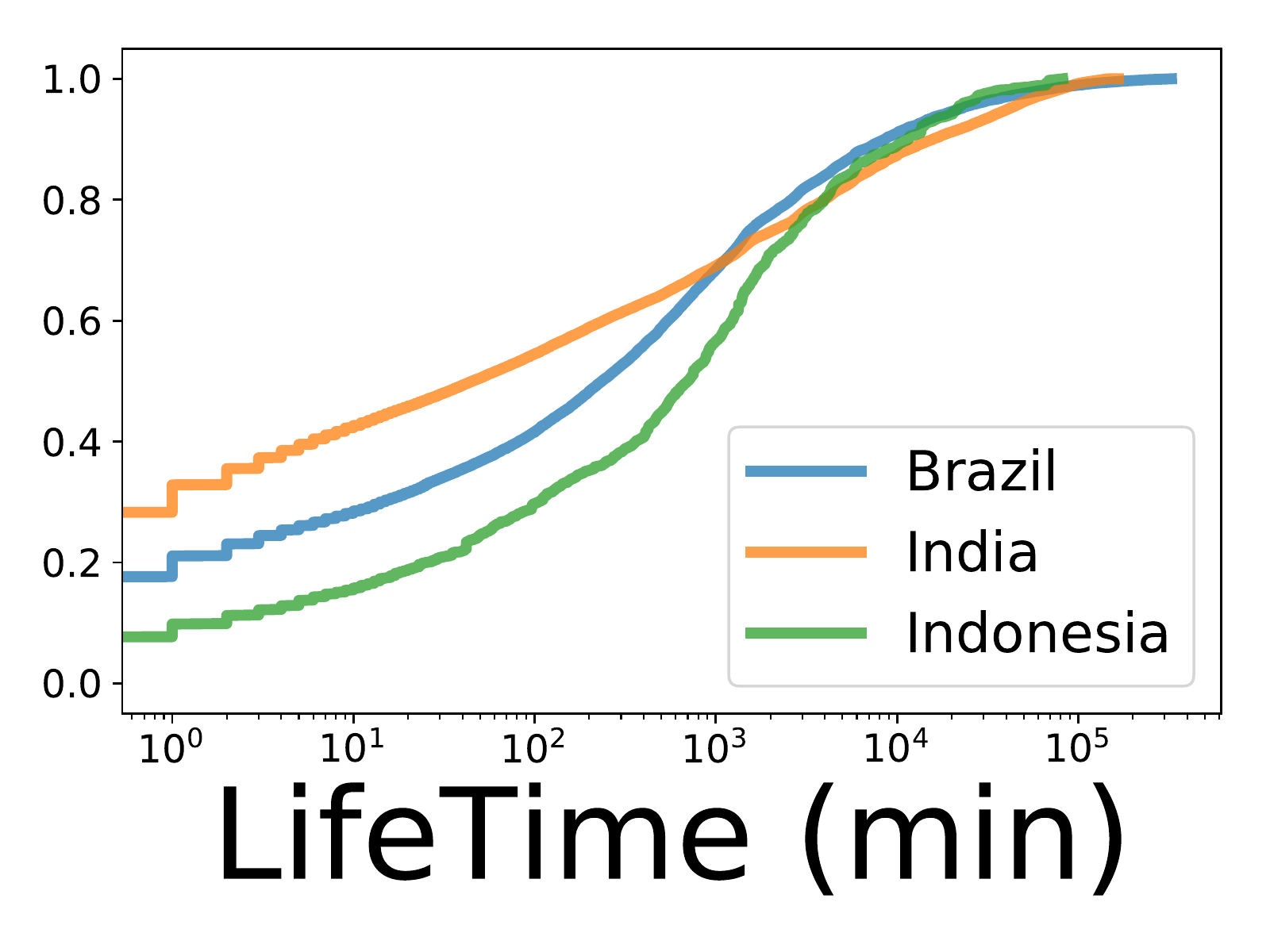}
		\caption{Lifetime}
		\label{fig:lifetime2}
	%\qquad
	\end{subfigure}%\hspace{-0.2cm}
	\begin{subfigure}[b]{0.24\textwidth}
		\includegraphics[width=\textwidth]{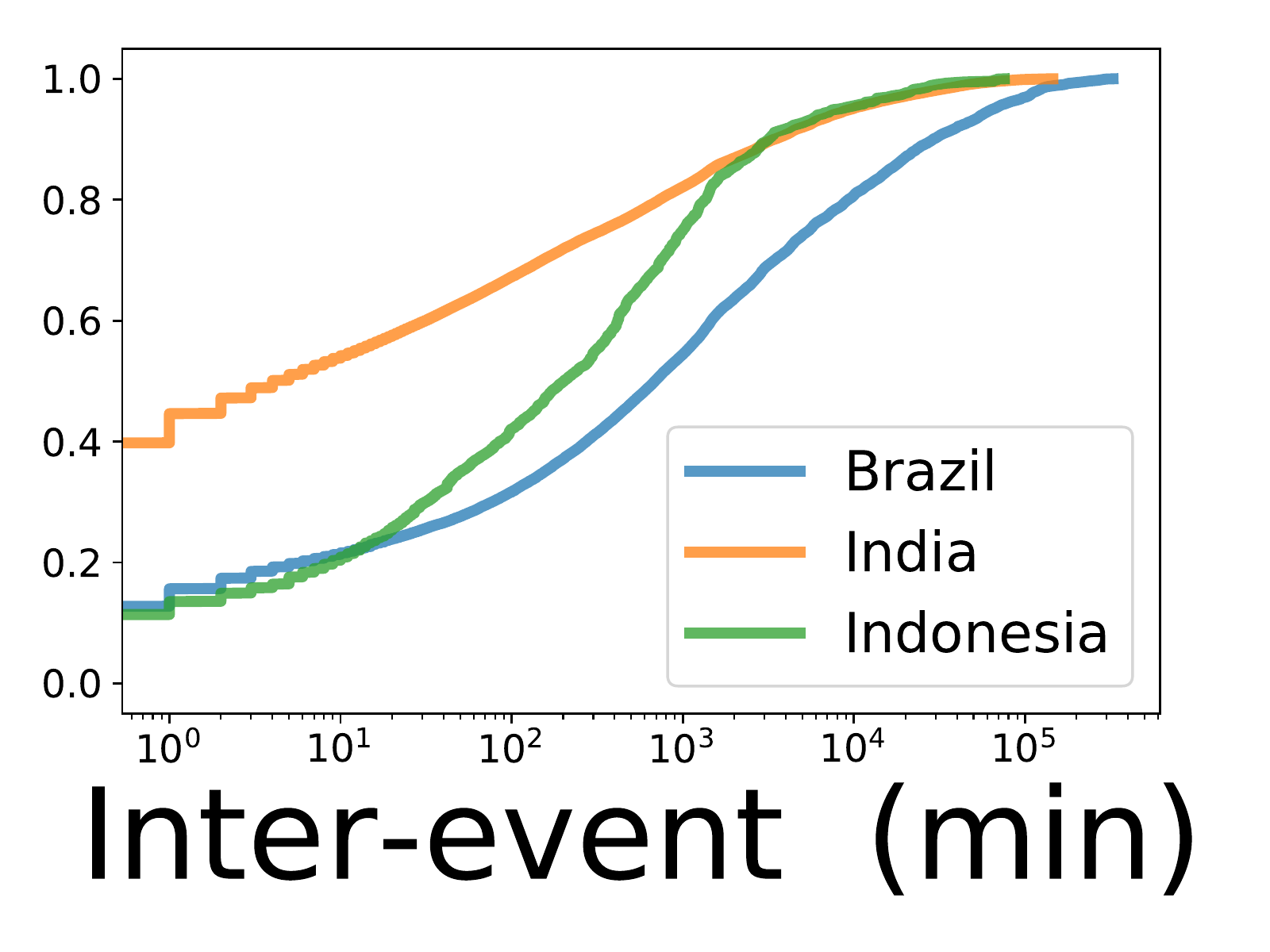}
		\caption{Inter-event}
		\label{fig:burstime2}
	%\qquad
	\end{subfigure}%\hspace{-0.2cm}
	\vspace{-0.2cm}
    \caption{CDF of sharing coverage and time dynamics metrics of images shared at least twice on WhatsApp.}
	\label{fig:times2}
	\vspace{-0.5cm}
\end{figure}

First, we calculate the total number of shares of each image and how many groups they have appeared. Figures \ref{fig:totalshares},\ref{fig:totalgroups} show the Cumulative Distribution Function (CDF) of the total number of shares and the number of distinct groups each image appeared in. 
Even though nearly 80\% images on WhatsApp were posted only once, there are some very popular images broadly shared over 100 times that reached multiple groups. This shows that WhatsApp can be used as a mass communication media and the potential of virality of content.

\textbf{Time Analysis for WhatsApp Data:}
Besides looking at the spread of images on WhatsApp, we also analyze their ``lifetimes'' in Figure~\ref{fig:lifetime2}. 
The lifetime is given by the difference between the last and first occurrence of the image in our dataset. In short, while most of the images (80\%) last no more than 2 days, there are images in Brazil and in India that continued to appear even after 2 months of the first appearance ($10^5$ minutes). 
We can also see that the majority (60\%) of the images are posted before 1000 minutes after their first appearance. Moreover, in Brazil and India, around 40\% of the shares were done after a day of their first appearance and 20\% after a week. 
Further analysis, in Figure~\ref{fig:burstime2} shows the distribution of the ``inter-event times'' between posts of the same image. We observe that the inter-event time of images in India is much faster than in Brazil and Indonesia, i.e., more than 50\% of posts are done in intervals of 10 minutes or less, while just 20\% of shares were done in this same time interval in Brazil and Indonesia. We manually looked for reasons behind the short period of time between posts and found that in the data from India, there is more automated, spam-like behavior compared to in Brazil and Indonesia.

In conclusion, these results suggest that WhatsApp is a very dynamic network and most of its image content is ephemeral, i.e., the images usually appear and vanish quickly. The linear structure of chats make it difficult for an old content to be revisited, yet there are some that linger on the network longer, disseminating over weeks or even months.

\section{Network structure}
\label{sec:networks}
\vspace{-0.1cm}

In this section, we investigate the network structure of public WhatsApp groups and compare its characteristics with other real  and synthetic social networks.
To create a network from the WhatsApp groups, we connected two groups if they share a common user.
%, as Reddit, Flickr, and with synthetic graphs. 
Although WhatsApp is an encrypted personal chat application, the possibility to create public groups allows multiple and socially distant users to connect to each other across the network, forming a complex social structure able to flow high volumes of information. 
Although the WhatsApp group network resembles many other social networks, little is known about the differences in information dissemination. In this section, we investigate how the structure of WhatsApp groups and users differ from other networks by using traditional complex network metrics. 

\begin{figure}[h]
	\centering
	%\hspace{-0.2cm}
	\begin{subfigure}[c]{0.25\textwidth}
		\includegraphics[trim=0cm 0.35cm 0cm 0cm, clip=True, width=\textwidth]{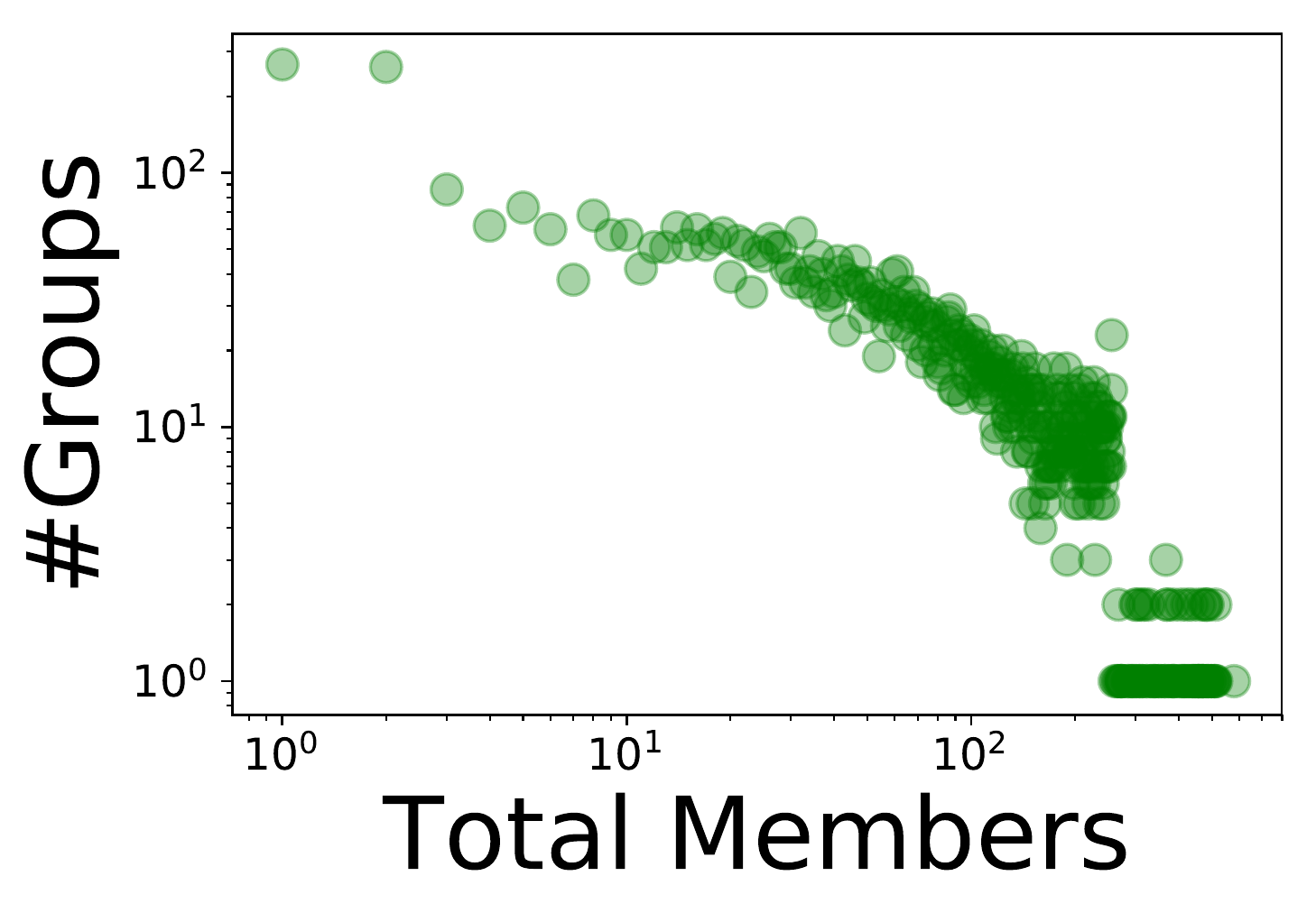}
		\caption{Wpp. India}
		\label{fig:ug_india}
	\end{subfigure}%\hspace{-0.3cm}
	\begin{subfigure}[c]{0.245\textwidth}
		\includegraphics[trim=0cm 0cm 0cm 0.35cm, clip=True, width=\textwidth]{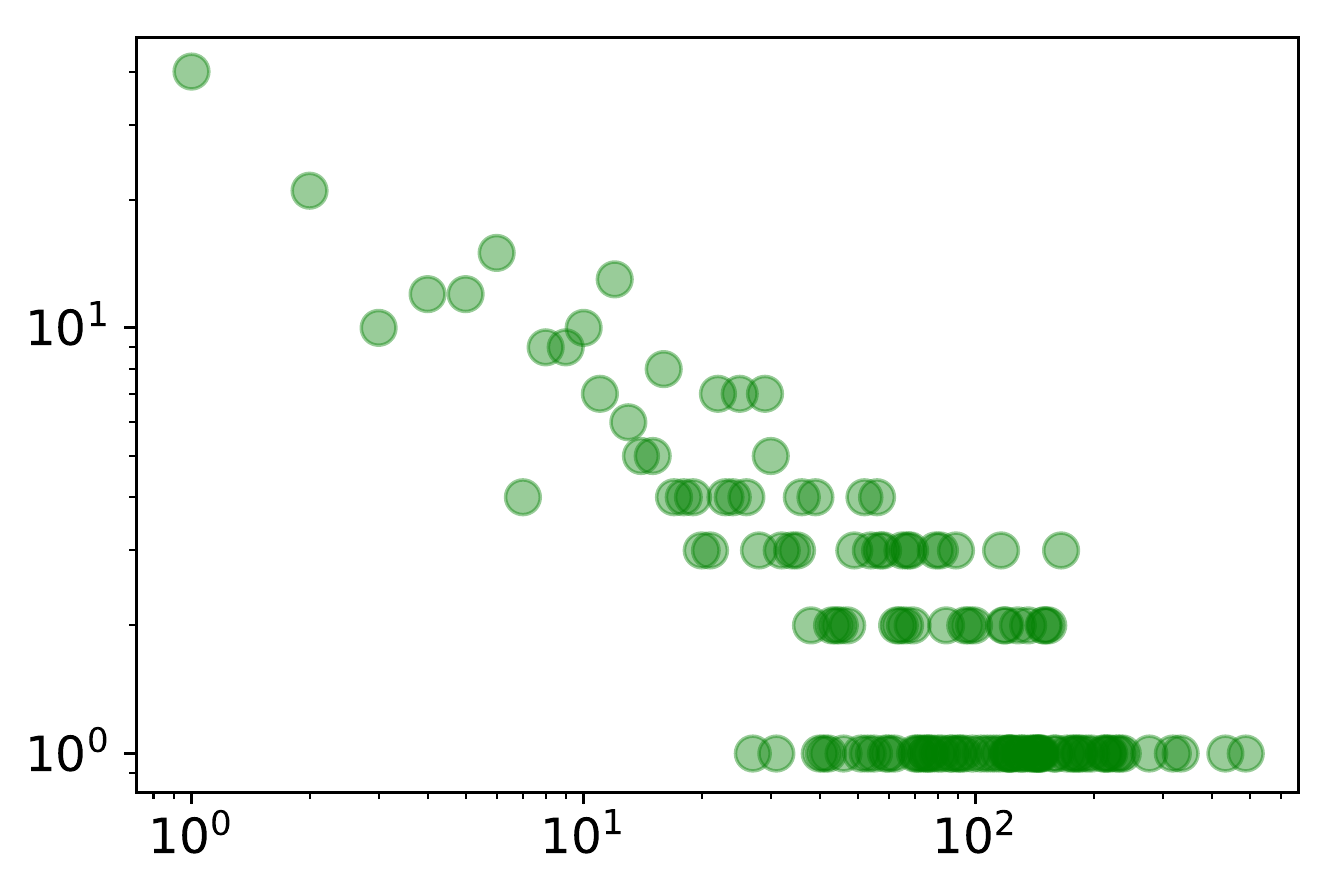}
		\caption{Wpp. Brazil}
		\label{fig:ug_brazil}
	\end{subfigure}%\hspace{-0.3cm}
	\begin{subfigure}[c]{0.245\textwidth}
		\includegraphics[trim=0cm 0cm 0cm 0.35cm, clip=True, width=\textwidth]{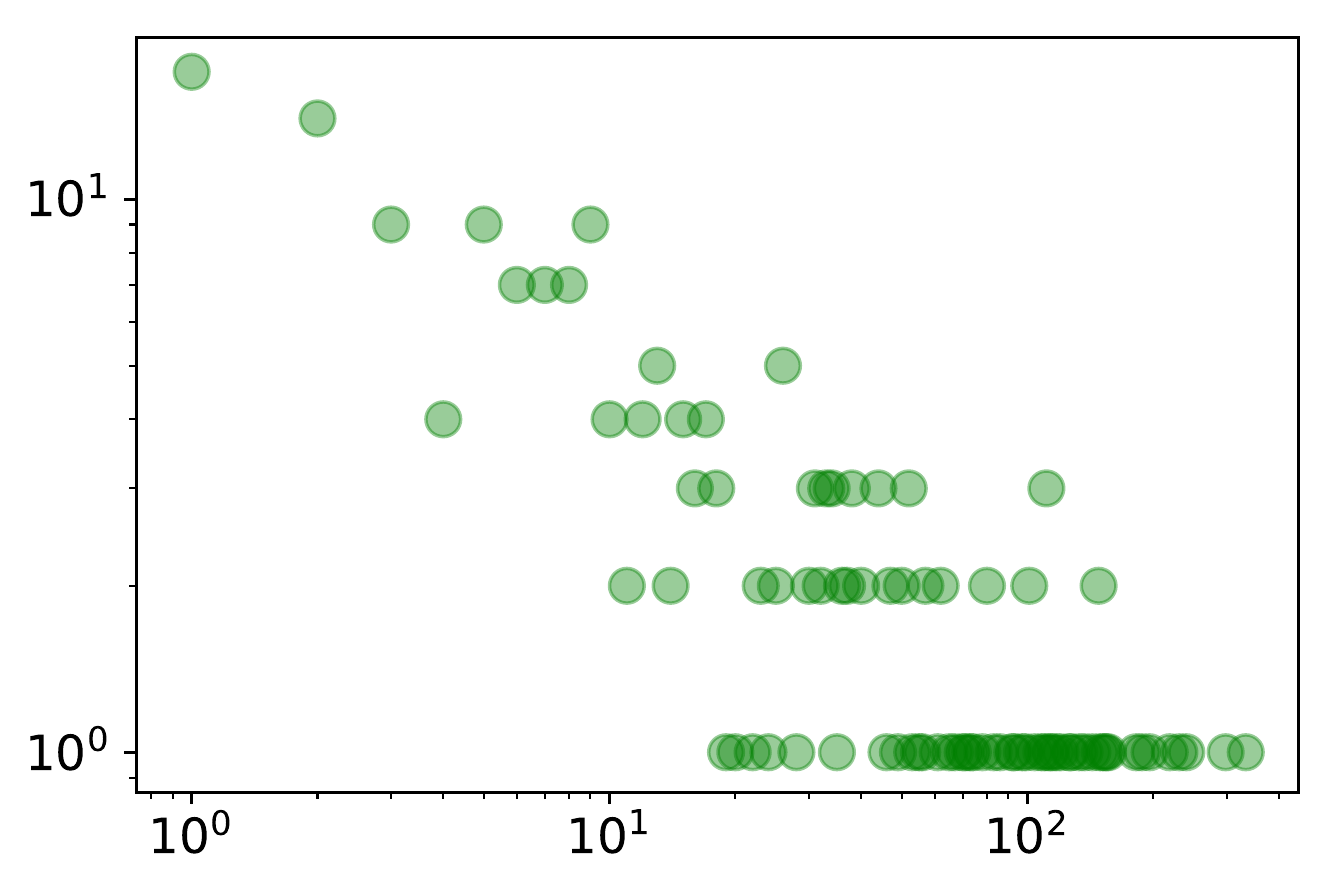}
		\caption{Wpp. Indonesia}
		\label{fig:ug_indonesia}
	\end{subfigure}%\hspace{-0.3cm}
	\begin{subfigure}[c]{0.245\textwidth}
		\includegraphics[trim=0cm 0cm 0cm 0.35cm, clip=True, width=\textwidth]{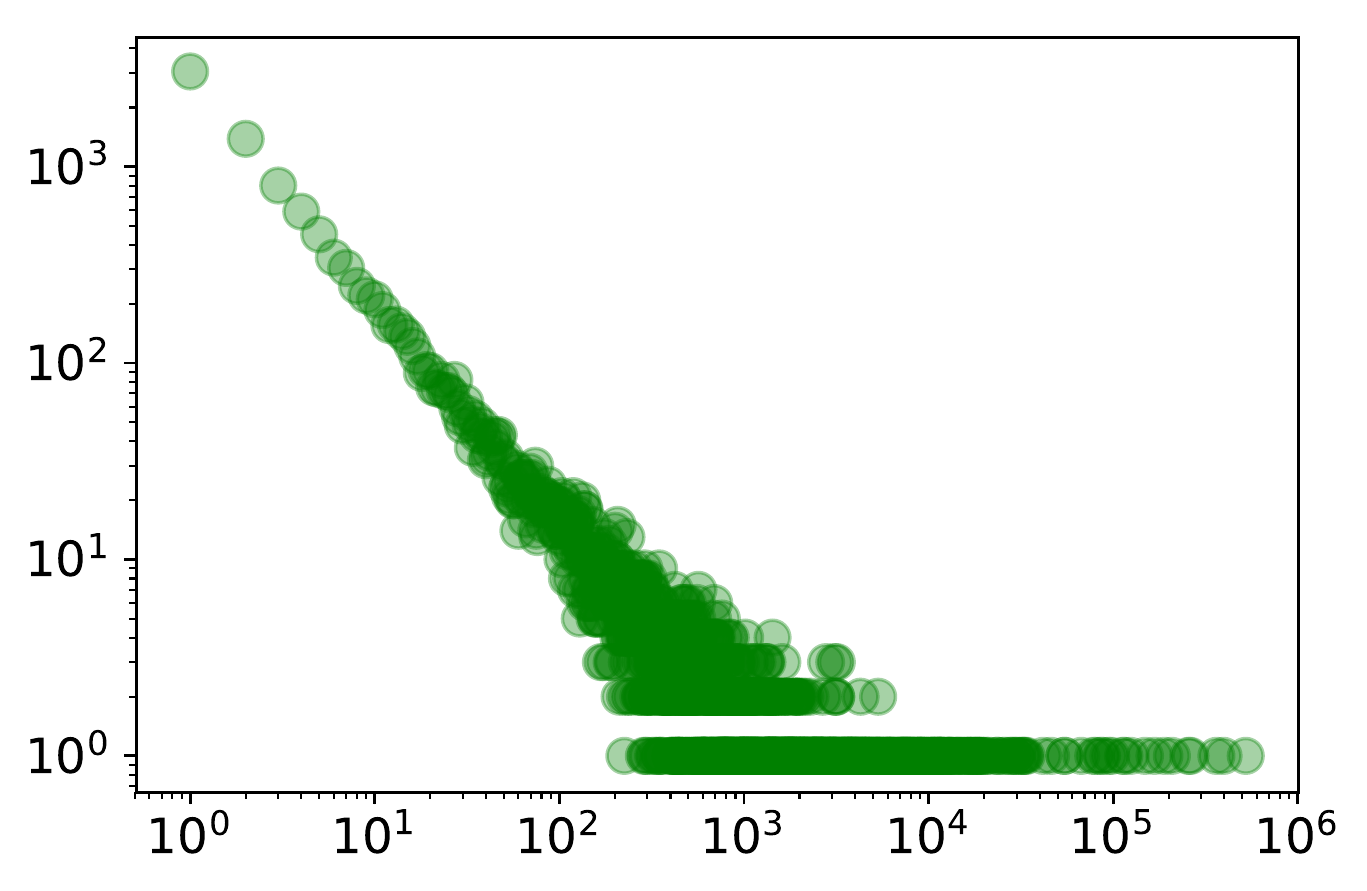}
		\caption{Reddit}
		\label{fig:ug_reddit}
	\end{subfigure}
	
		%\hspace{-0.2cm}
	\begin{subfigure}[c]{0.25\textwidth}
		\includegraphics[trim=0cm 0.35cm 0cm 0cm, clip=True, width=\textwidth]{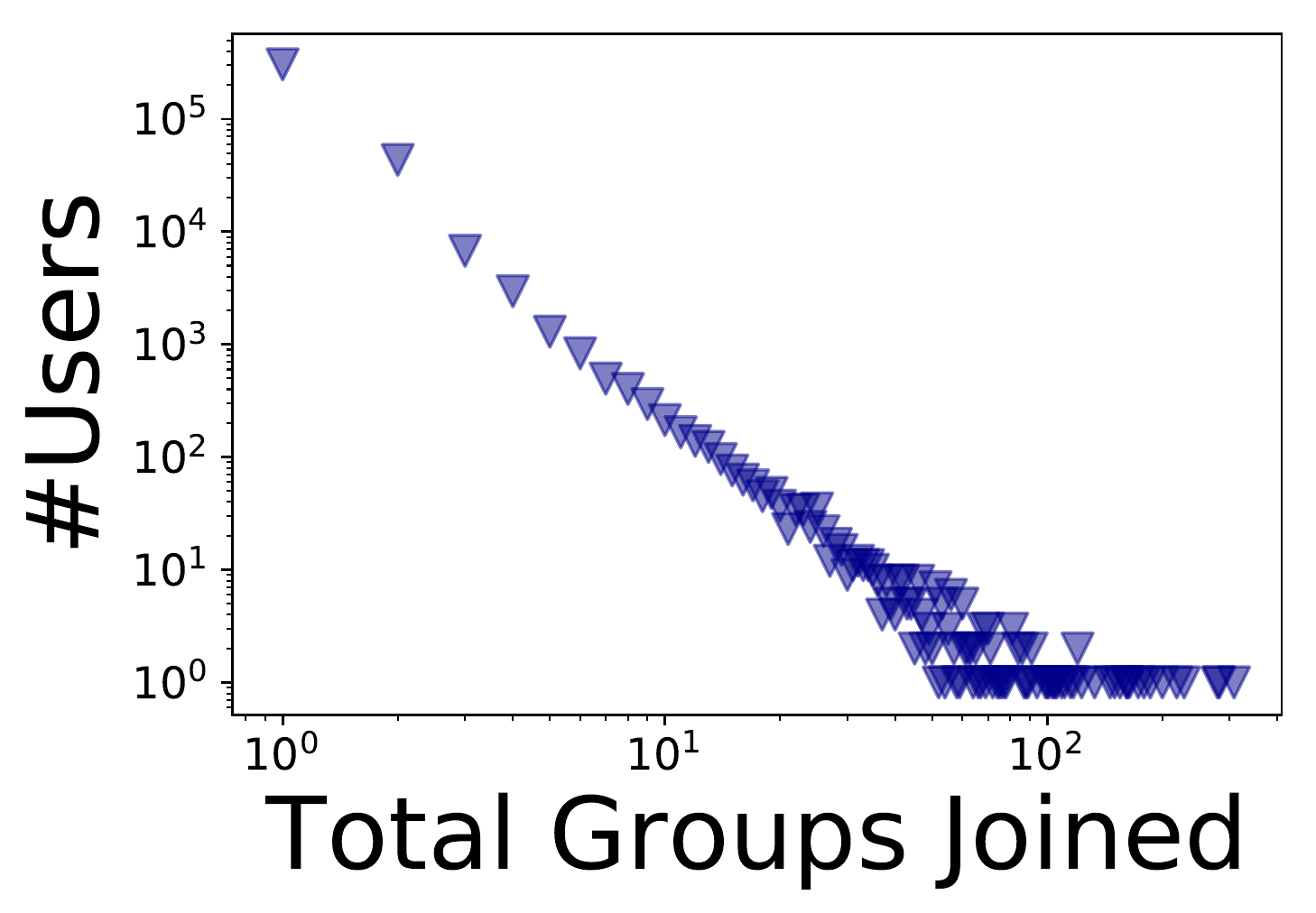}
		\caption{Wpp. India}
		\label{fig:gu_india}
	\end{subfigure}%\hspace{-0.3cm}
	\begin{subfigure}[c]{0.245\textwidth}
		\includegraphics[trim=0cm 0cm 0cm 0.35cm, clip=True, width=\textwidth]{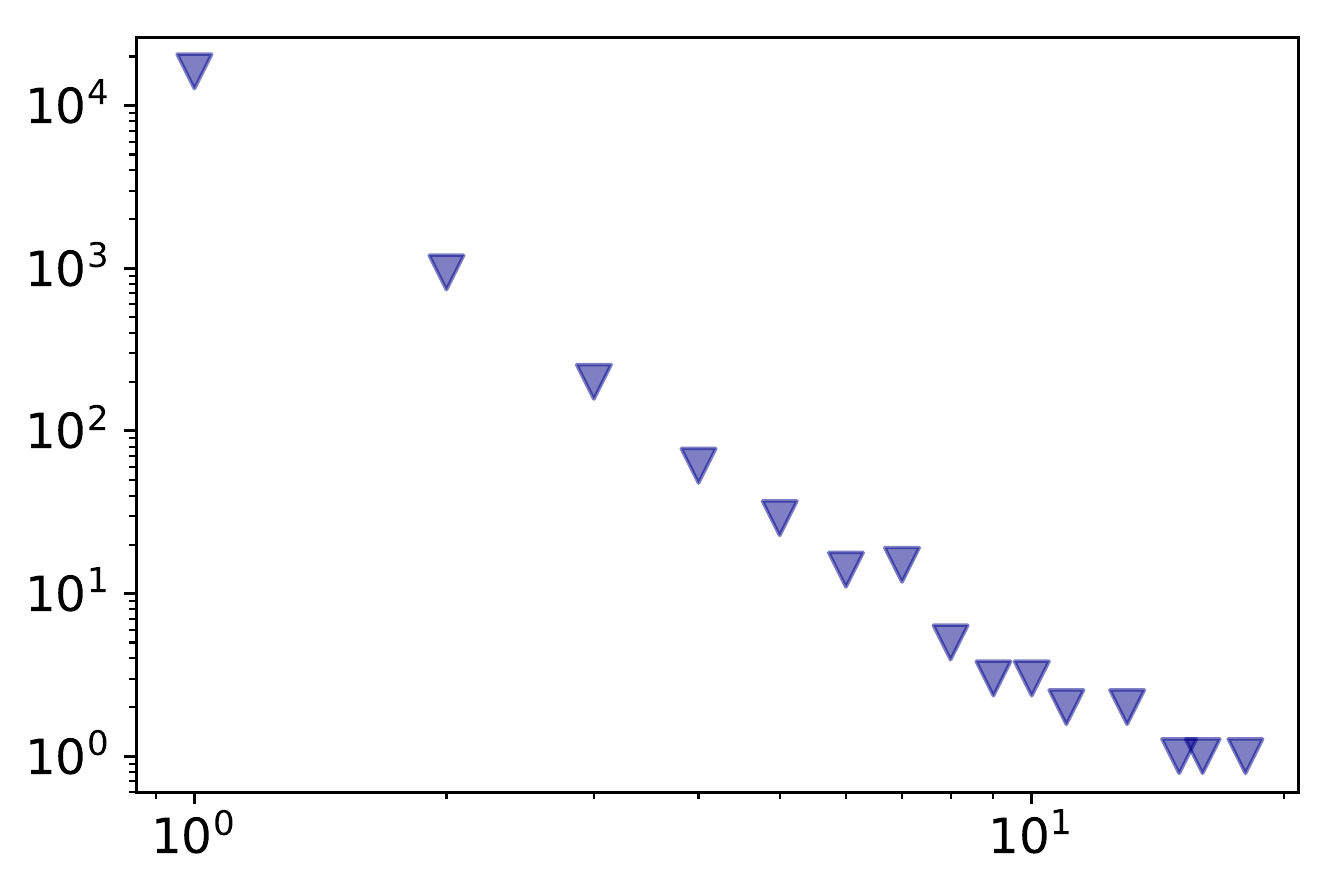}
		\caption{Wpp. Brazil}
		\label{fig:gu_brazil}
	\end{subfigure}%\hspace{-0.3cm}
	\begin{subfigure}[c]{0.245\textwidth}
		\includegraphics[trim=0cm 0cm 0cm 0.35cm, clip=True, width=\textwidth]{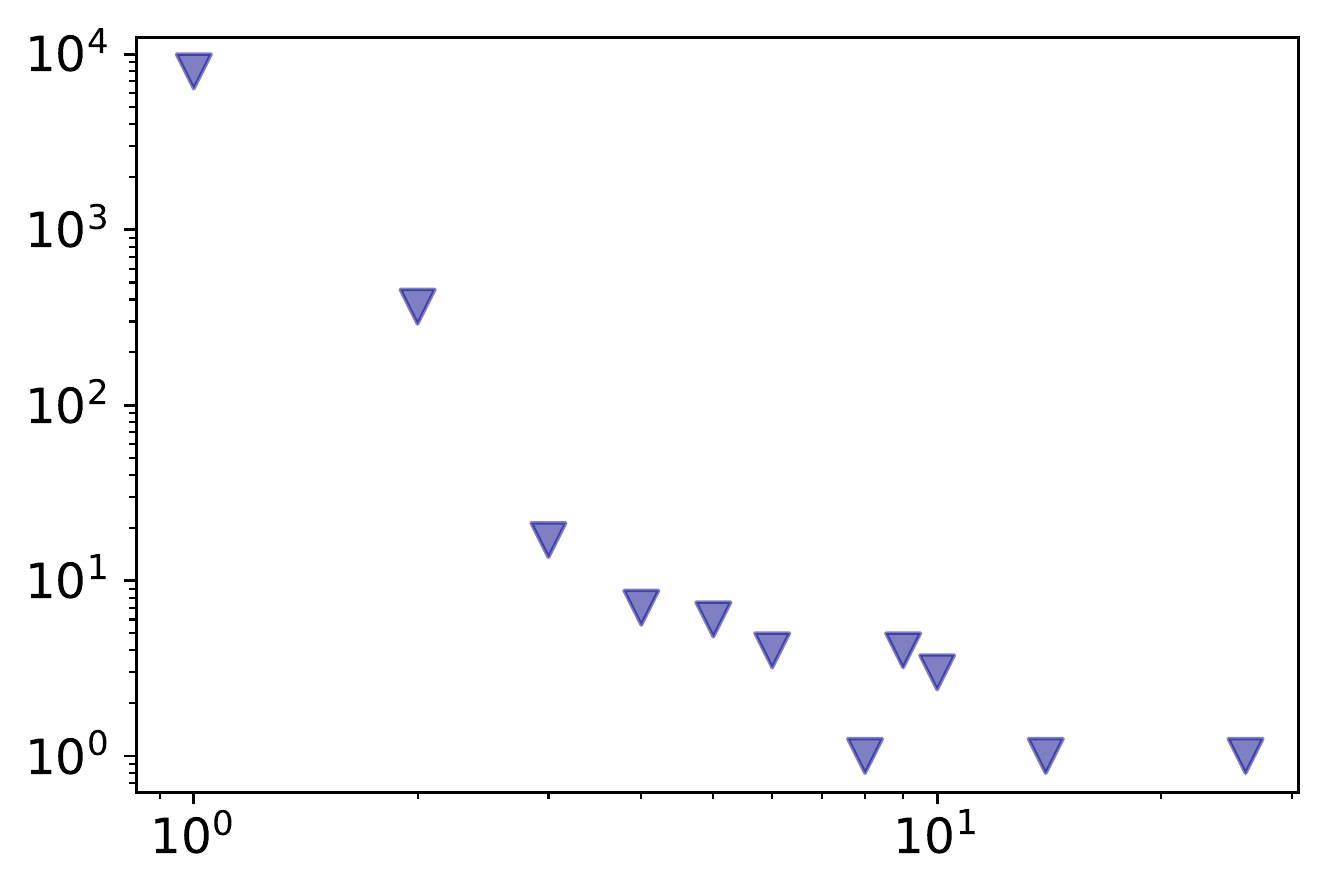}
		\caption{Wpp. Indonesia}
		\label{fig:gu_indonesia}
	\end{subfigure}%\hspace{-0.3cm}
	\begin{subfigure}[c]{0.245\textwidth}
		\includegraphics[trim=0cm 0cm 0cm 0.35cm, clip=True, width=\textwidth]{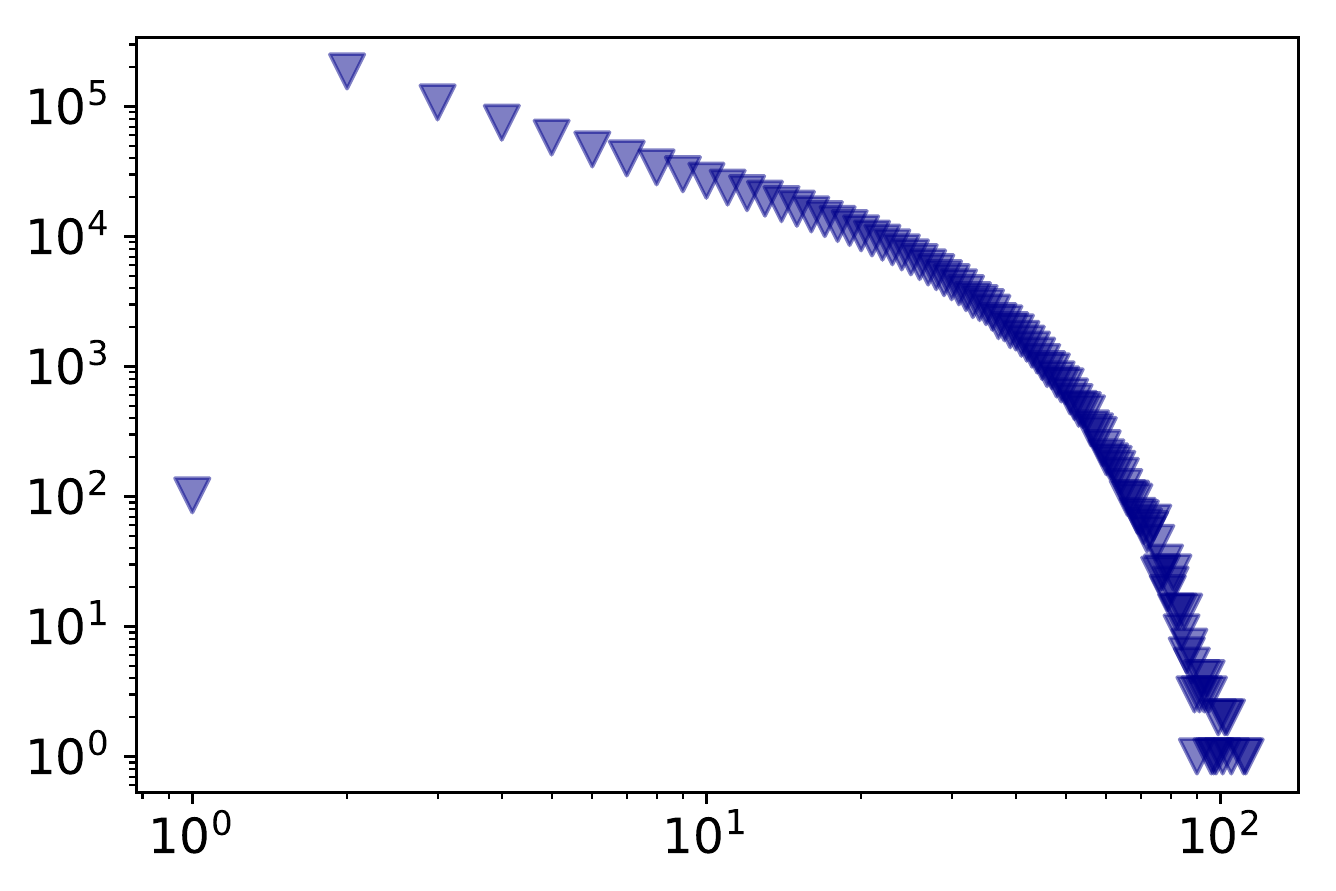}
		\caption{Reddit}
		\label{fig:gu_reddit}
	\end{subfigure}
	
	\vspace{-0.2cm}
	\caption{Distributions of the number of members per group and total groups joined per user in WhatsApp (Wpp) and in Reddit.}
	\label{fig:usergroups}
	\vspace{-\baselineskip}
%	\vspace{-0.3cm}
\end{figure}

In Figure~\ref{fig:usergroups}, we show the distribution of groups per user and users per group. We can compare these characteristics with Reddit, as subreddits can be viewed as groups. Observe that the maximum of 256 members in groups is a determining element in the network, capable of limiting group size, mainly in India (Figure \ref{fig:ug_india}), where there are over 300k users and more than 5k groups.\footnote{In our data, some groups have more than 256 members, because our data is a temporal snapshot and members can leave join groups during this time.} 
On the other hand, in Reddit, where there is no limit, it is possible to see that the group size can be as large as $10^5$ members, what creates big hubs of users. As both platforms have no limit on the number of groups users can join, we expected to see no differences in the total number of groups users participate. However, note that in Reddit, the distribution has a exponential decay, with a limit on $\approx 100$ groups. On the other hand, all WhatsApp curves are similar with a well behaved power law curve, which naturally yields a larger variance. Note that in India we have users who participated in more than $300$ groups.

\begin{figure}[t]
	\centering \hspace{-0.1cm}
	\begin{subfigure}[c]{0.31\textwidth}
		\includegraphics[width=\textwidth]{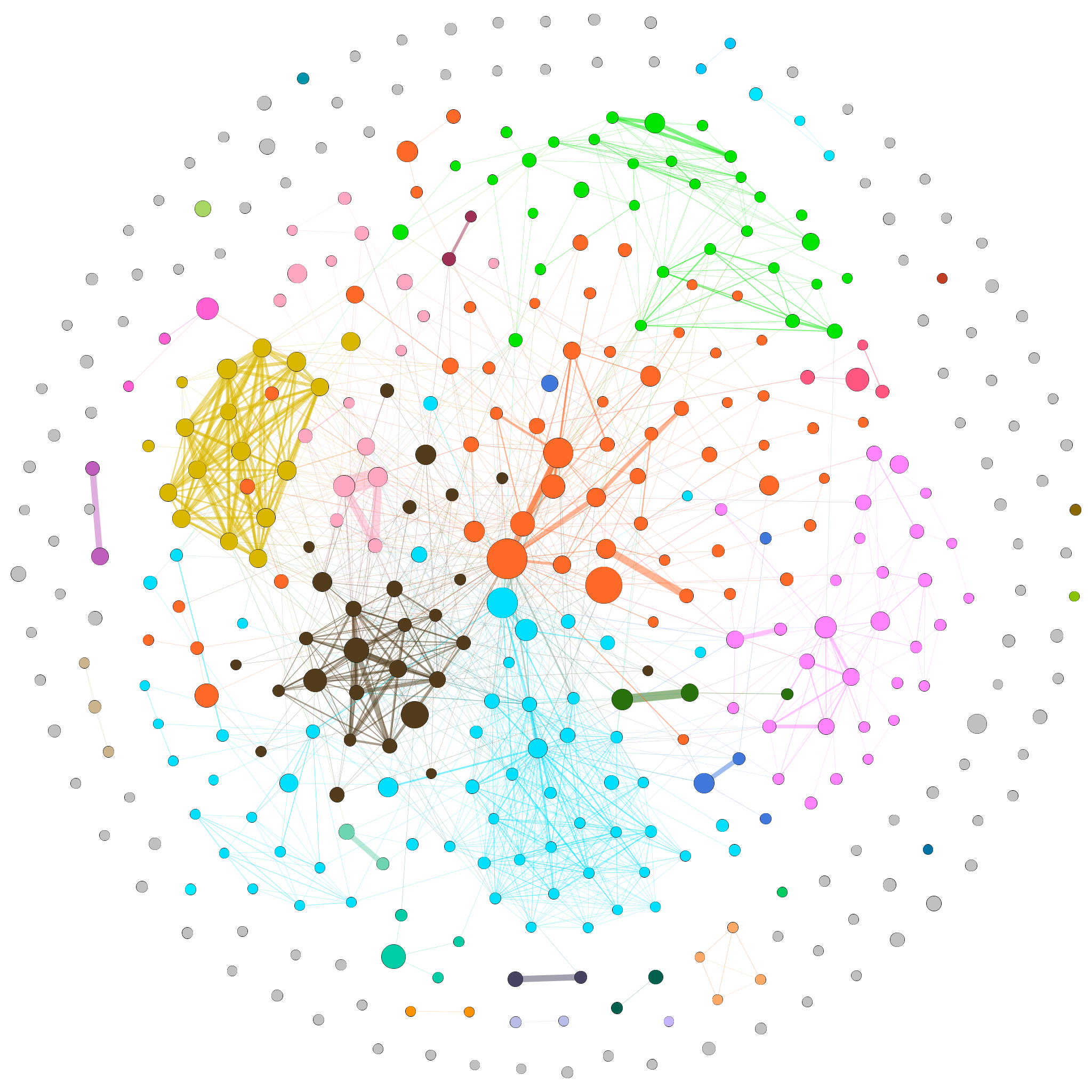}
		\caption{Brazil}
		\label{fig:network_brazil}
	%\qquad
	\end{subfigure}
	%\vspace{-0.1cm}
	\begin{subfigure}[c]{0.32\textwidth}
		\includegraphics[width=\textwidth]{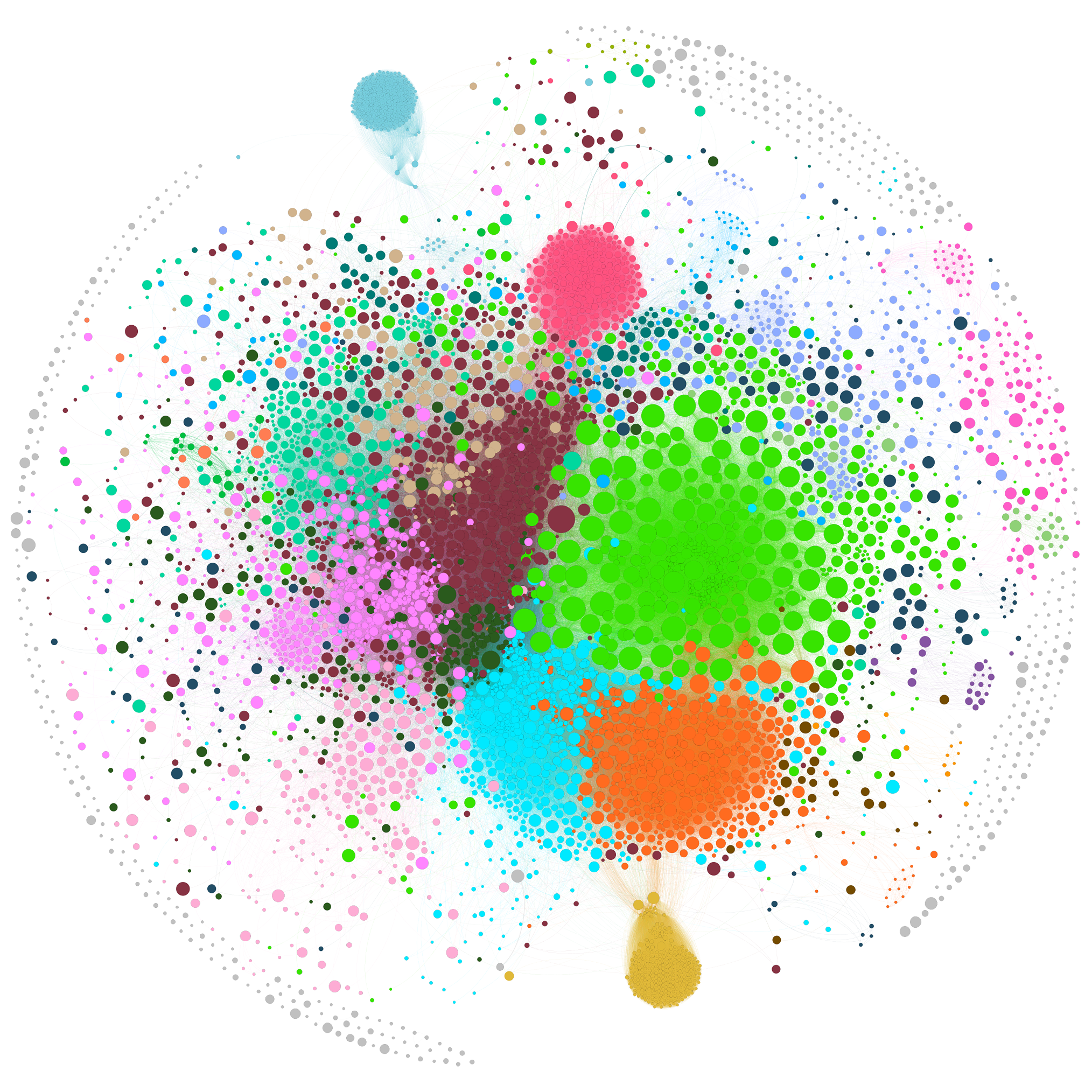}
		\caption{India}
		\label{fig:network_india}
	\end{subfigure}
	%\vspace{-0.1cm}
	\begin{subfigure}[c]{0.31\textwidth}
		\includegraphics[width=\textwidth]{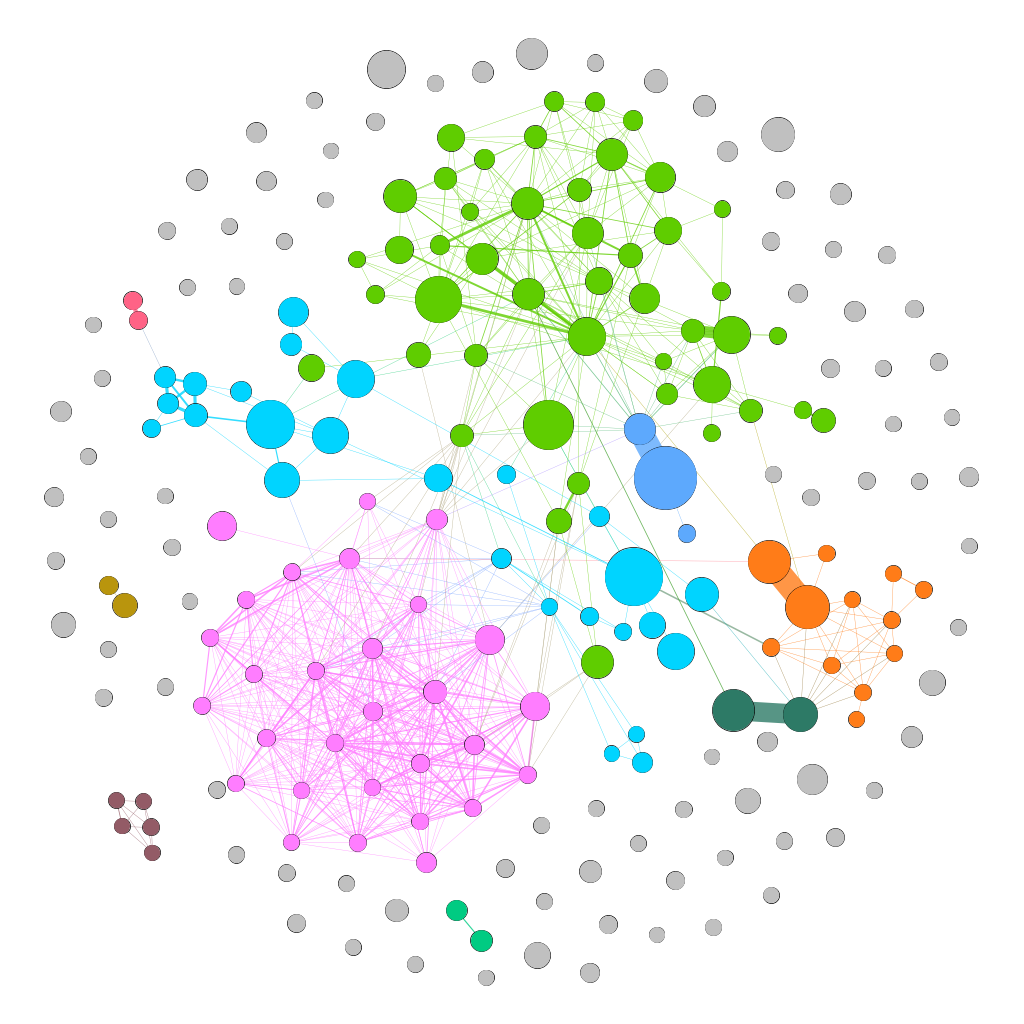}
		\caption{Indonesia}
		\label{fig:network_indonesia}
	\end{subfigure}
	
	\vspace{-0.2cm}
	\caption{WhatsApp Public Groups Network for each country. Each node is a group and edges represents members in common.}
	\label{fig:networks}
	\vspace{-0.55cm}
\end{figure}

In Figure~\ref{fig:networks} we show these networks for all three countries. The size of the node is proportional to the number of members in that group. We colored nodes according to its community in that graph following the modularity  algorithm~\cite{blondel2008fast}. Observe that in all graphs there is an evident largest connected component and some other group clusters. Also, note that some groups position themselves as bridges and hubs, connecting different communities of the network structure.

Next, we compare the  characteristics of the WhatsApp group network and other social network graphs:
(i) random generated graphs using the Barabasi-Albert scale free model, the Erdős–Rényi model, the small world model~\cite{watts1998collective} and the Forest Fire network model~\cite{leskovec2005graphs}, for which we used the same number of nodes in the Indian dataset in order to create a comparable network; 
(ii) the network of subreddits from Reddit~\cite{olson2015navigating}, and, 
(iii) the Flickr network~\cite{mcauley2012image}, which, different from the WhatsApp and Reddit group networks, the Flickr graph represents the network of images shared by users on the platform.
The results are shown in Table~\ref{tab:networks}.
We observe that WhatsApp shares common characteristics with other real-world social networks: high clustering coefficient, %where nodes tend to create tightly knit clusters. 
 giant largest connected component, and small average path length, which are all typical properties of a social network.
The only aberration is the slightly higher diameter than others graphs analyzed.
WhatsApp also shows a higher Pearson coefficient, in which nodes tend to be connected with other nodes with similar degree values. 
In epidemic analyses, it can help to understand the spreading of infection across the network, as a misinformation campaign targeting high degree groups is likely to spread to other high degree nodes.

\begin{table}[t]
\scriptsize
\center
\caption{Network metrics for the public groups network from WhatsApp compared to other networks.}
%\vspace{-0.5cm}
\begin{tabular}{l|l|l|l|l|l|l|l|l|l|}
\cline{2-10}
 & \multicolumn{1}{c|}{\#Nodes} & \multicolumn{1}{c|}{\#Edges} & \multicolumn{1}{c|}{\begin{tabular}[c]{@{}c@{}}Mean \\ Degree\end{tabular}} & \multicolumn{1}{c|}{\begin{tabular}[c]{@{}c@{}}Clustering\\ Coefficient\end{tabular}} & \multicolumn{1}{c|}{Diameter} & \multicolumn{1}{c|}{APL$^*$} & \multicolumn{1}{c|}{Density} & \multicolumn{1}{c|}{LCC$^{**}$} & \multicolumn{1}{c|}{\begin{tabular}[c]{@{}c@{}}Pearson\\ Coefficient\end{tabular}} \\ \hline
\multicolumn{1}{|l|}{Wapp. India} & 5,839 & 407,081 & 69.71 & 0.59 & 11 & 3.17 & 0.0239 & 92.6\% & 0.295 \\ \hline
\multicolumn{1}{|l|}{Wapp. Brazil} & 414 & 1,400 & 6.76 & 0.32 & 8 & 3.19 & 0.0164 & 65.2\% & 0.346 \\ \hline
\multicolumn{1}{|l|}{Wapp. Indonesia} & 217 & 699 & 6.44 & 0.38 & 9 & 3.09 & 0.0298 & 55.3\% & 0.290 \\ \hline
\multicolumn{1}{|l|}{Bar.-Albert} & 5,839 & 792,300 & 271.38 & 0.10 & 3 & 1.95 & 0.0465 & 100\% & 0.008 \\ \hline
\multicolumn{1}{|l|}{Erdos-Renyi} & 5,839 & 1,534,952 & 525.76 & 0.09 & 2 & 1.91 & 0.0901 & 100\% & -0.001 \\ \hline
\multicolumn{1}{|l|}{Smallworld} & 5,839 & 604,250 & 206.97 & 0.34 & 3 & 1.98 & 0.0355 & 100\% & 0.007 \\ \hline
\multicolumn{1}{|l|}{ForestFire} & 5,839 & 12,930 & 4.43 & 0.42 & 17 & 5.25 & 0.0008 & 100\% & -0.066 \\ \hline
\multicolumn{1}{|l|}{Reddit} & 15,122 & 4,520,054 & 597.81 & 0.82 & 6 & 2.03 & 0.0395 & 99,8\% & -0.045 \\ \hline
\multicolumn{1}{|l|}{Flickr} & 105,938 & 2,316,948 & 43.74 & 0.09 & 	9 & 	4.8 & 0.0004 & 99.8\% & 0.247 \\ \hline
\end{tabular}
\flushleft
\vspace{-0.15cm}
\scriptsize{*Average Path Length} \newline
\scriptsize{**Largest Connected Component}
\label{tab:networks}
\vspace{-0.5cm}
\end{table}

\section{Impact of forwarding limitations on information spread}
\label{sec:sei}
\vspace{-0.1cm}

We use the epidemiological model of Susceptible-Exposed-Infected (SEI)~\cite{guihua2004global} to estimate the virality of malicious messages in WhatsApp groups by assuming misinformation as an infection that spreads to users through the group network. In our scenario, the nodes are members of various groups and the infected nodes can spread the infection to a entire group at once, exposing all their participants. In this model, \textit{Susceptible} (S) is the initial condition in which the user did not have any contact with the infection; \textit{Exposed} (E) are those who received the misinformation through any of the groups they participate, but didn't share it; \textit{Infected} (I) is the final stage in which a user who was exposed to the content shares this message in the network.
This model has two basic parameters: \textit{virality} ($\alpha$) and \textit{exposition} ($\beta$). We also implemented a third parameter \textit{forward limit} ($\varphi$) to test the restrictions on sharing by WhatsApp.

The \textbf{virality} ($\alpha$) of malicious content is a parameter that controls the rate of infected users. 
This parameter indicates the probability of an exposed user to share the content that she had contact with. 
We consider that users are infected when they forward or broadcast this content, as it indicates a degree of belief in the shared message. 
The \textbf{exposition} parameter ($\beta$) refers to the rate at which exposed users become infected. It represents the probability of an exposed user to transform in an infected one.
Lastly, the \textbf{forward limit} ($\varphi$) of infection is a specific parameter we use to restrict the spread of the infection, to simulate the actual conditions on WhatsApp. This parameter indicates the maximum amount of groups an infected node can spread the infection to. 
We started our simulation by selecting one user randomly to be the initial infected node to start the spreading. % to begin the dissemination and expose other users. 
For each user exposed, they have a probability given by $\alpha$ to share the malicious message. When these infected nodes decide to forward, there is a limitation given by $\varphi$, the maximum number of groups they will send the content to. After that, each user in the groups that received the message are exposed. Then, each exposed user has also a probability $\beta$ of becoming an infected node and sharing the content.
We repeatedly iterate this process until the dissemination stops or when all users are infected. 

\textbf{Experimental Results.}
We perform several experiments using our SEI model comparing the dissemination in different scenarios by enforcing limits of broadcast and forward. %For those simulations we consider 
Since it would not be possible to reach isolated nodes using the whole structure, only the largest connected component was considered.

\begin{figure}[t]
	\centering \hspace{-0.2cm}
	\begin{subfigure}[c]{0.3\textwidth}
		\includegraphics[trim=0cm 0cm 0cm 1cm, clip=True, width=\textwidth]{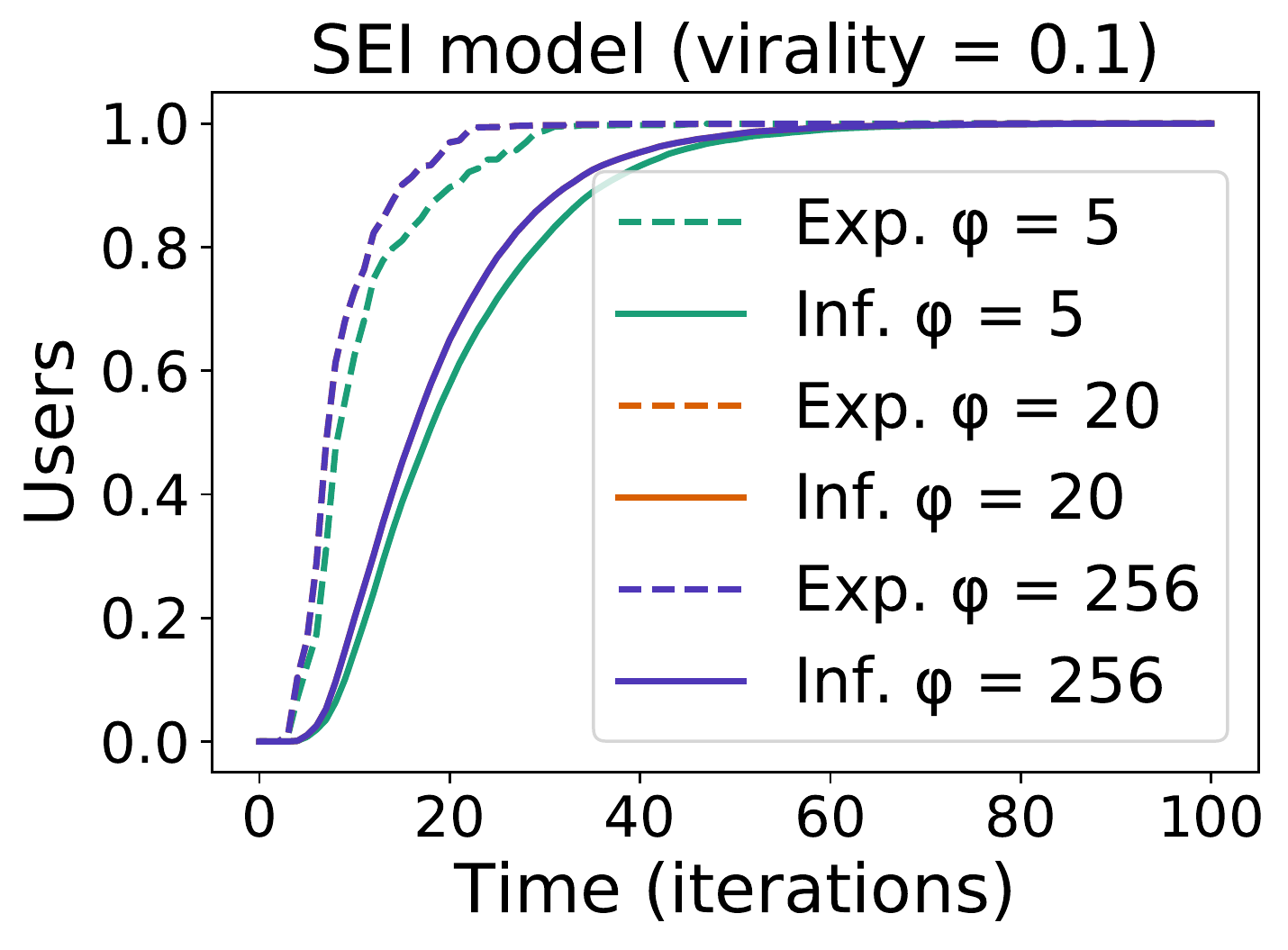}
		\caption{Brazil}
		\label{fig:sei-v01-brazil}
	\end{subfigure} \hspace{-0.2cm}
	\begin{subfigure}[c]{0.3\textwidth}
		\includegraphics[trim=0cm 0cm 0cm 1cm, clip=True, width=\textwidth]{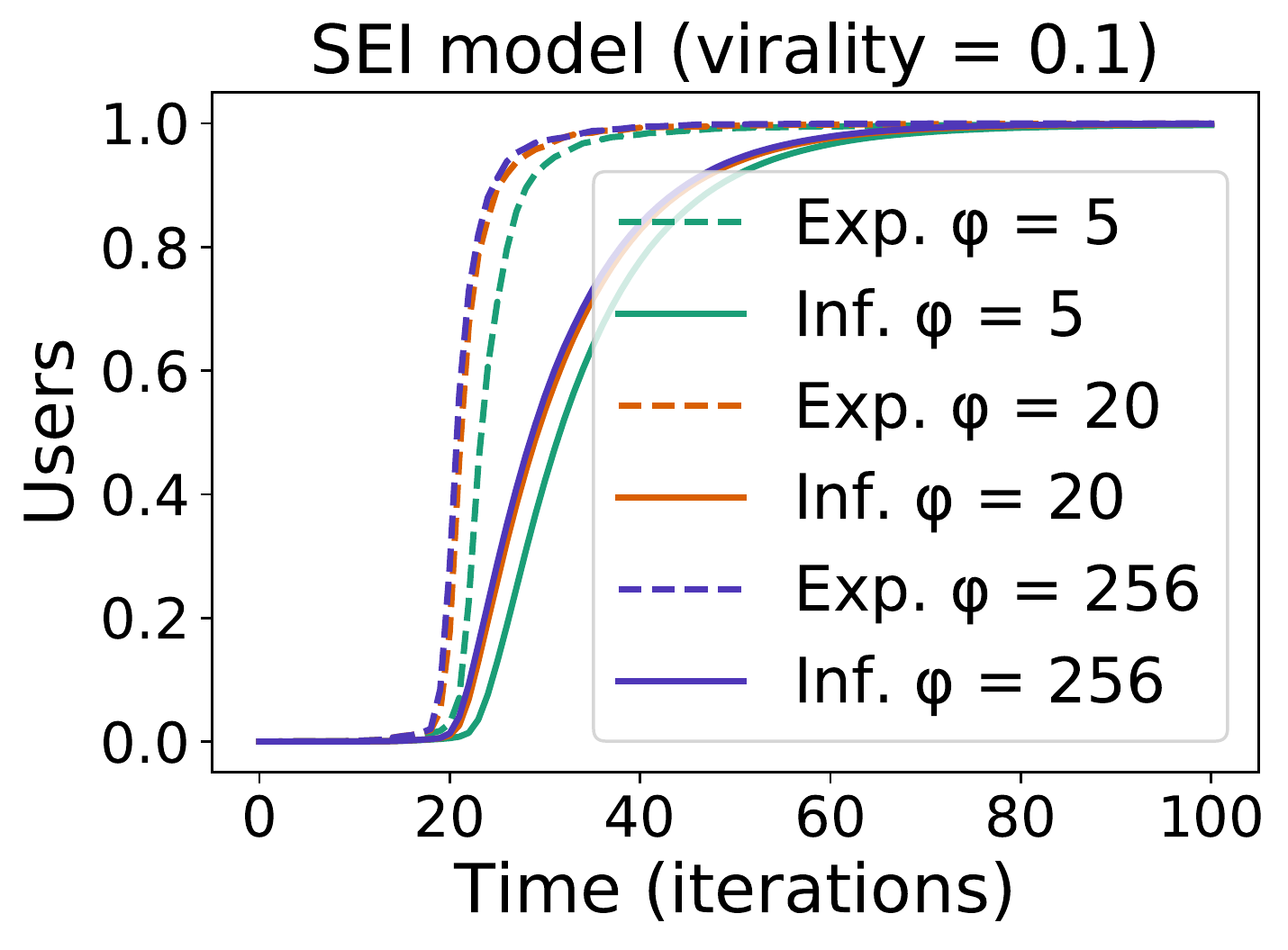}
		\caption{India}
		\label{fig:sei-v01-india}
	\end{subfigure} \hspace{-0.2cm}
	\begin{subfigure}[c]{0.3\textwidth}
		\includegraphics[trim=0cm 0cm 0cm 1cm, clip=True, width=\textwidth]{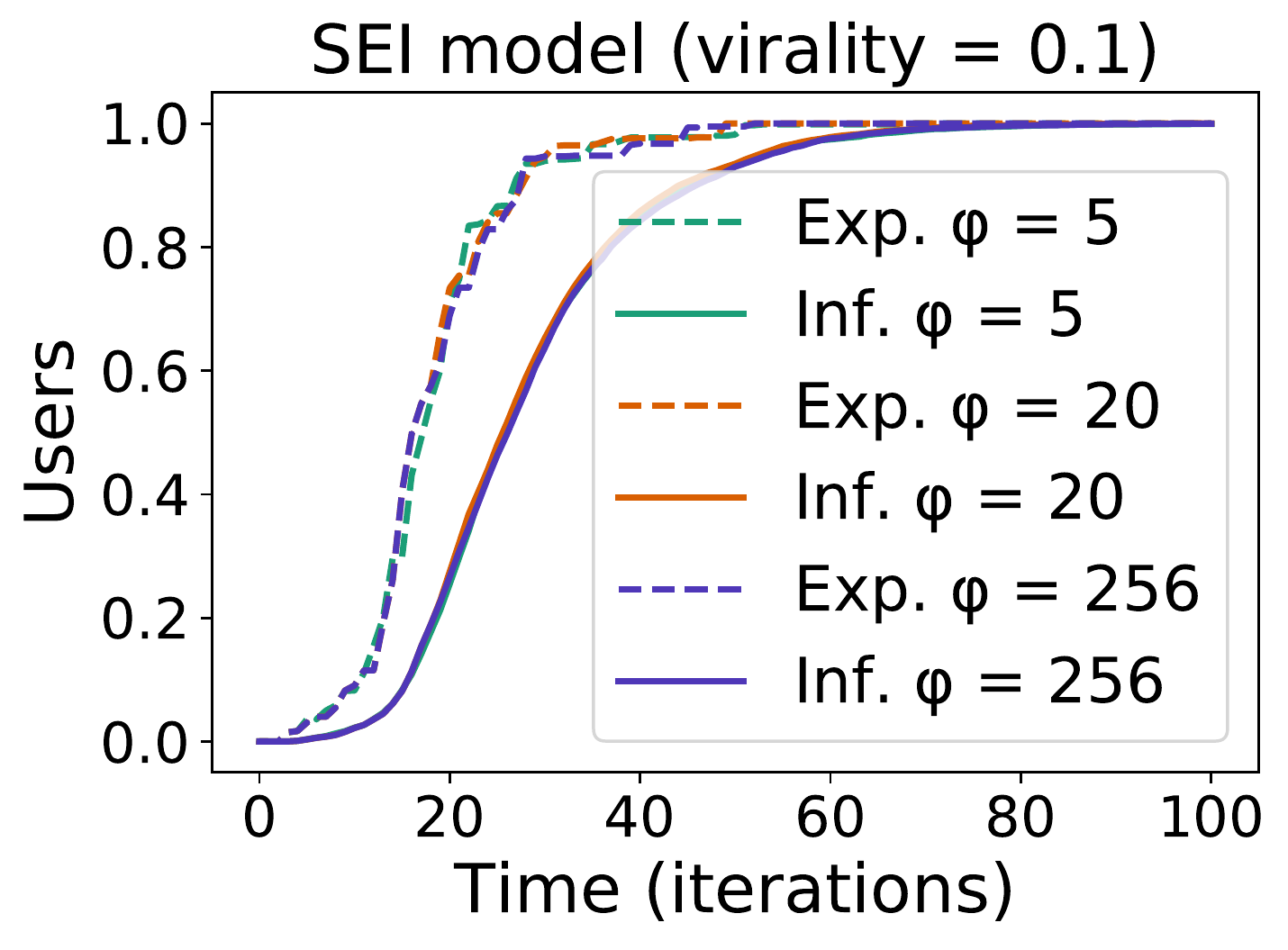}
		\caption{Indonesia}
		\label{fig:sei-v01-indonesia}
	\end{subfigure} %\hspace{-0.35cm}
	%\begin{subfigure}[Reddit]{\includegraphics[trim=0cm 0cm 0cm 1cm, clip=True, width=0.25\linewidth]{Figures/si-reddit-norm2-01-100.pdf}
	%\label{fig:sei-v01-reddit}}
	%\end{subfigure}
	\vspace{-0.3cm}
	\caption{SEI model %simulations restricted by groups 
	varying the forward limit $(\varphi)$. $\alpha = \beta = 0.1$.}
	\label{fig:sei-v01}
	\vspace{-0.1cm}
\end{figure}

Figure \ref{fig:sei-v01} shows the fraction of users infected over time for all three WhatsApp networks when the forward limit ($\varphi$) is varied, i.e., how the restrictions implemented by WhatsApp can interfere with the spread. We considered the limit of forwarding to 5 groups (the actual scenario), 20 groups (the previous limit), and 256 groups (the current limit for broadcasting). Notice that the rate of users exposed in the network grows very fast, regardless of forwarding limits, showing that a message can infect the entire network in 60 iterations. Also, observe that limitations on forwarding slightly diminish the velocity of spreading, but does not stop it completely, especially for exposed users.

\begin{figure}[t]
	\centering \hspace{-0.2cm}
	\begin{subfigure}[c]{0.3\textwidth}
		\includegraphics[trim=0cm 0cm 0cm 1cm, clip=True, width=\textwidth]{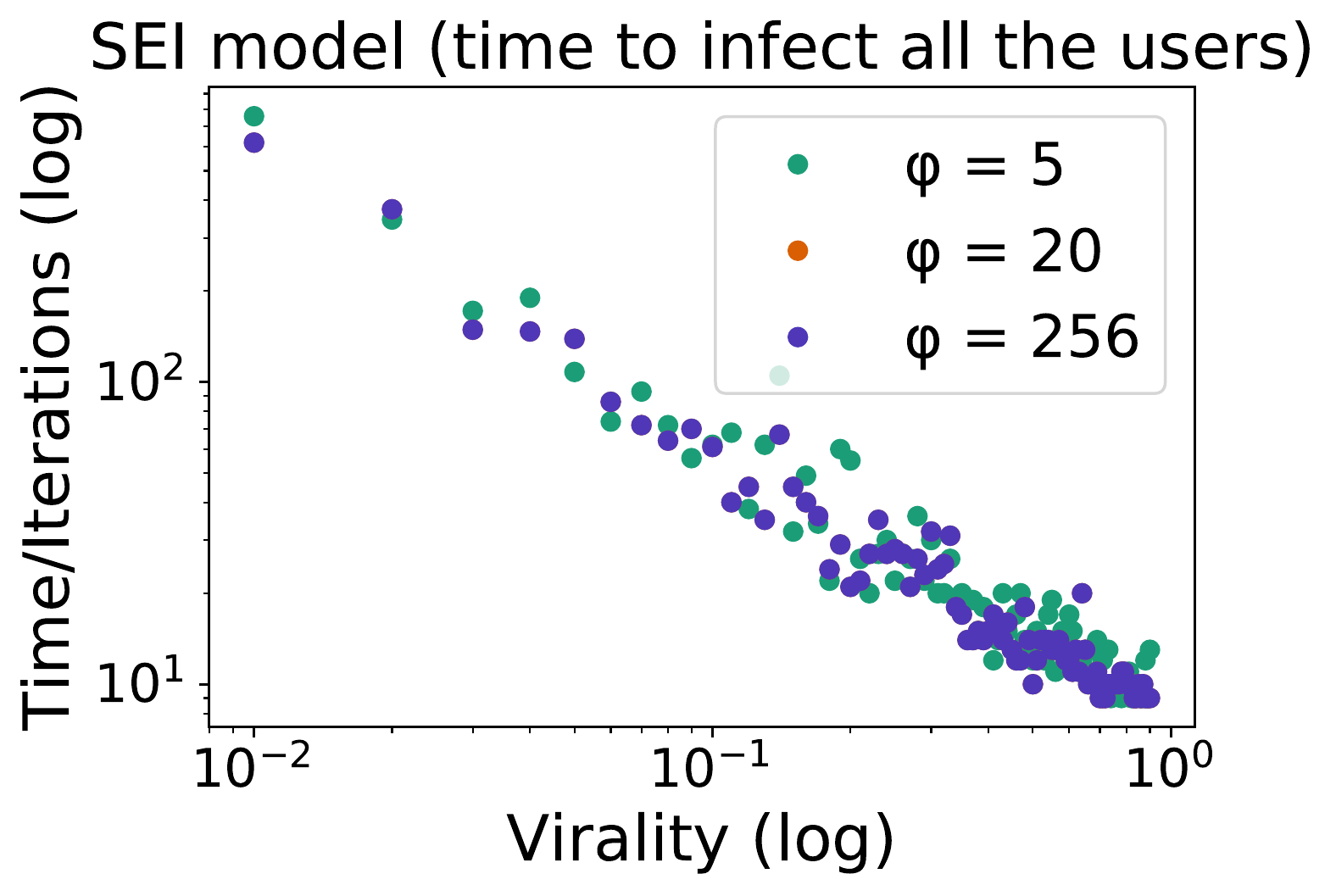}
		\caption{Brazil}
		\label{fig:sei_virality_brazil}
	\end{subfigure}\hspace{-0.2cm}
	\begin{subfigure}[c]{0.3\textwidth}
		\includegraphics[trim=0cm 0cm 0cm 1cm, clip=True, width=\textwidth]{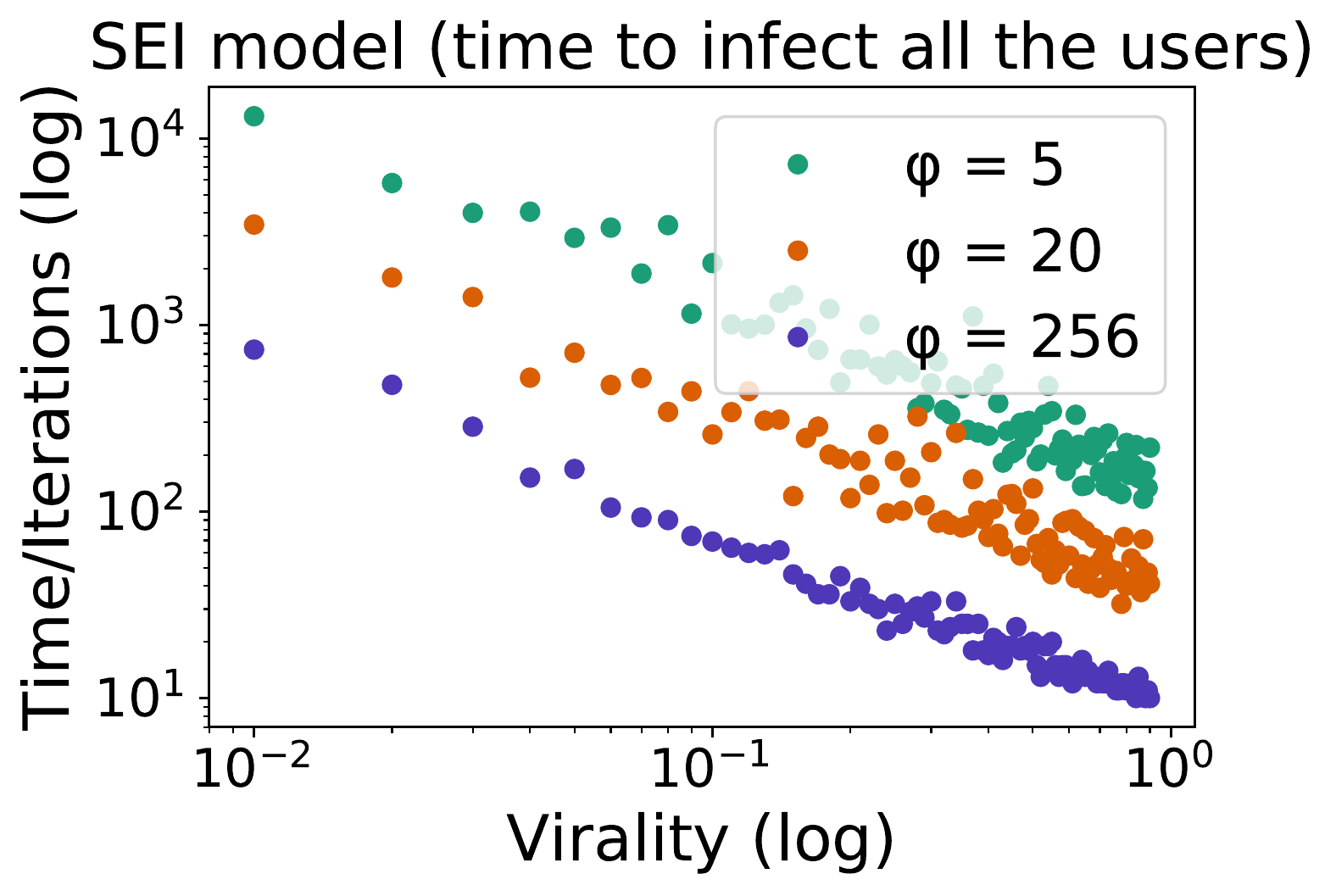}
		\caption{India}
		\label{fig:sei_virality_india}
	\end{subfigure}\hspace{-0.2cm}
	\begin{subfigure}[c]{0.3\textwidth}
		\includegraphics[trim=0cm 0cm 0cm 1cm, clip=True, width=\textwidth]{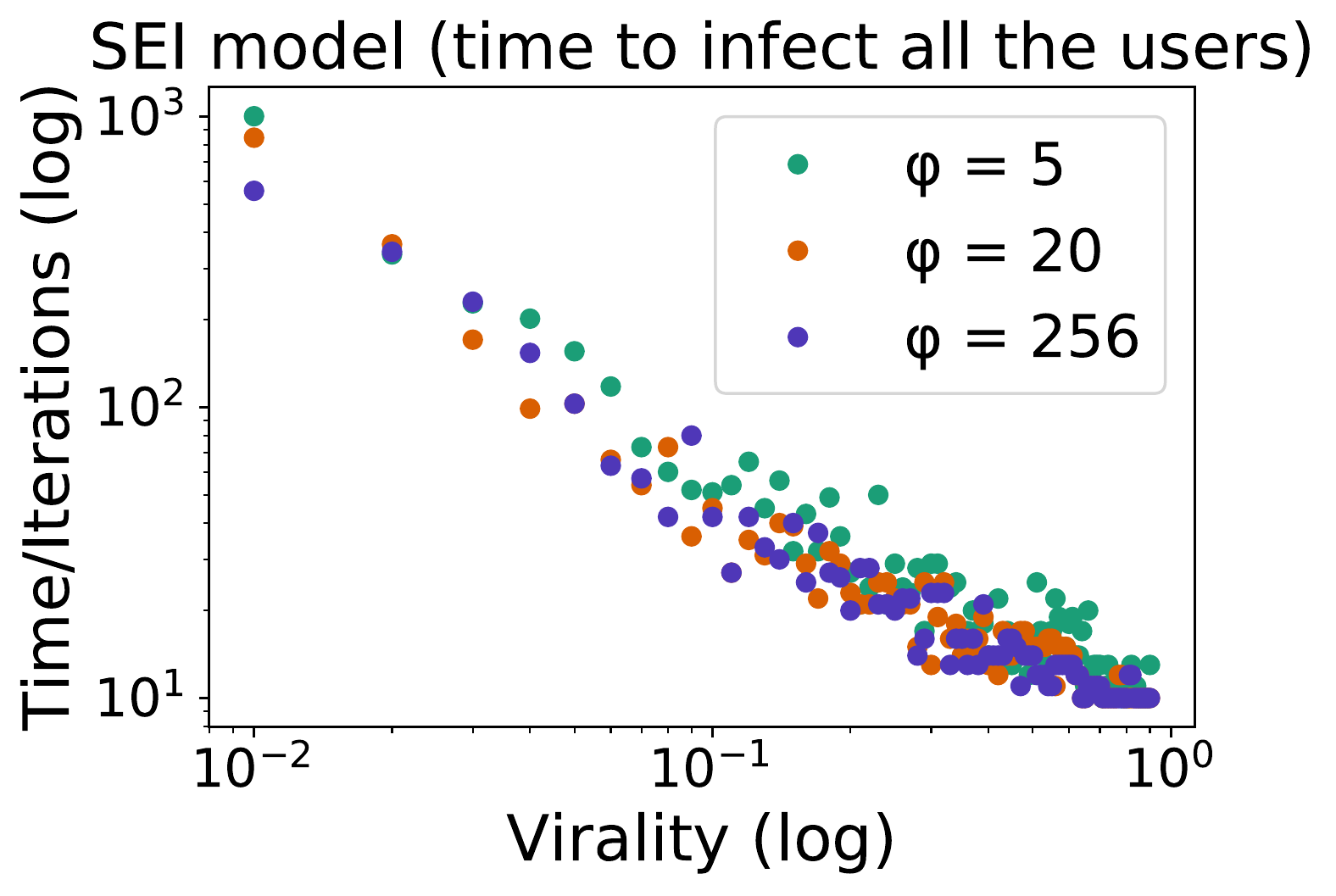}
		\caption{Indonesia}
		\label{fig:sei_virality_indonesia}
	\end{subfigure}
	\vspace{-0.2cm}
	\caption{Time to infect all user in the network on simulations of SEI model varying the virality $(\alpha)$ from $0.001$ up to $1.0$ .}
	\label{fig:sei_virality}
	\vspace{-0.55cm}
\end{figure}

We also evaluate the time needed for (mis)information with different potential viralities to infect all users.
Figure \ref{fig:sei_virality} shows the time needed to infect 100\% of the users by varying $\alpha$ from $0.001$ up to $1.0$, with different forwarding limits. Observe that in situations of mass dissemination (high $\alpha$), it is difficult to stop the infection because of the strong connections between groups. However, note that the limits in forwarding and broadcasting help to slow the propagation, mainly in larger networks, as in India. In short, \textbf{limits on forwarding and broadcasting can reduce velocity of dissemination by one order of magnitude for any of $\alpha$ virality}.

\begin{figure}[t]
	\centering \hspace{-0.2cm}
	\begin{subfigure}[c]{0.29\textwidth}
		\includegraphics[width=\textwidth]{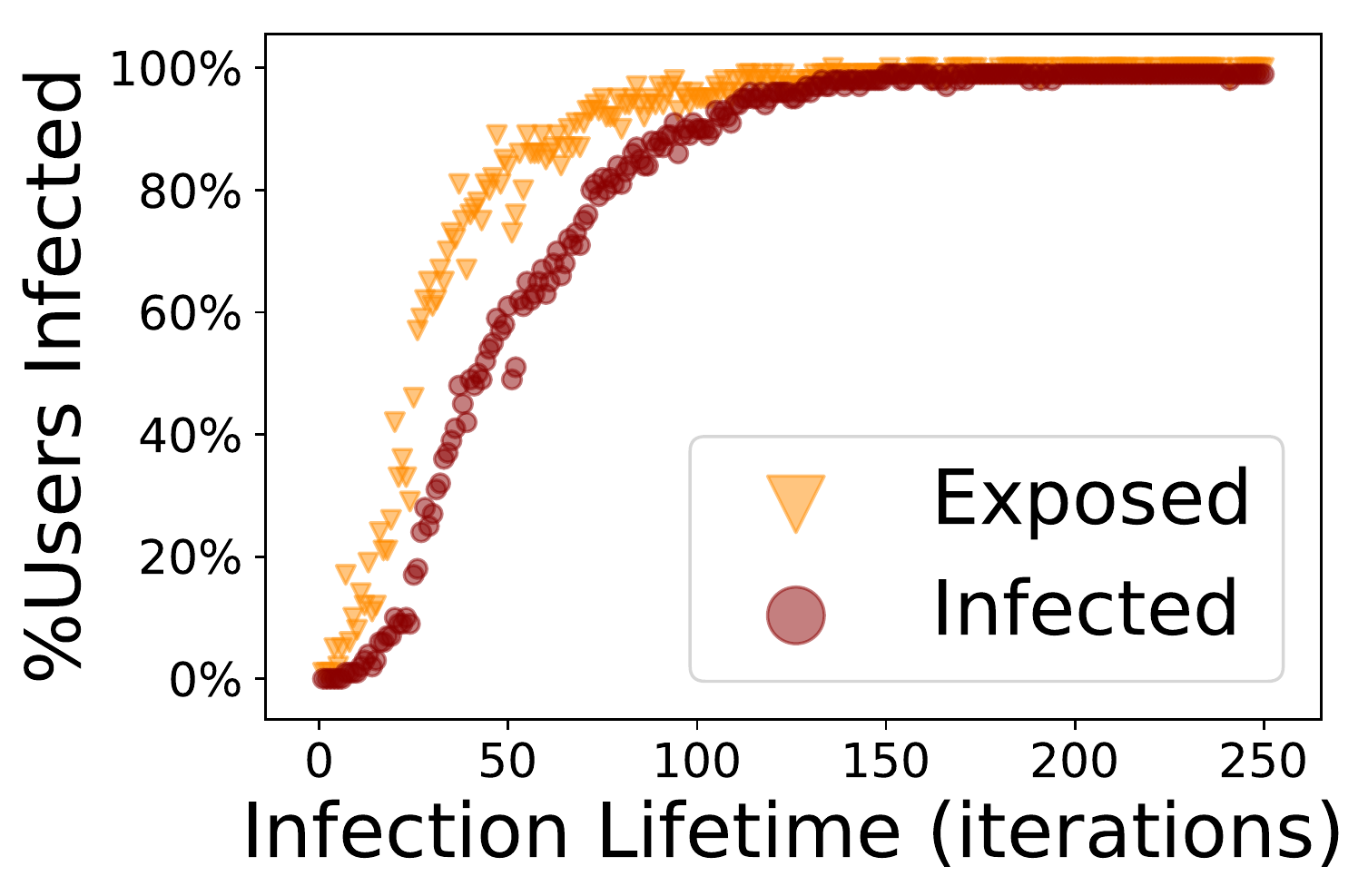}
		\vspace{-0.5cm}
		\caption{Brazil}
		\label{fig:sei_decay_brazil}
	\end{subfigure}\hspace{-0.15cm}
	\begin{subfigure}[c]{0.28\textwidth}
		\includegraphics[width=\textwidth]{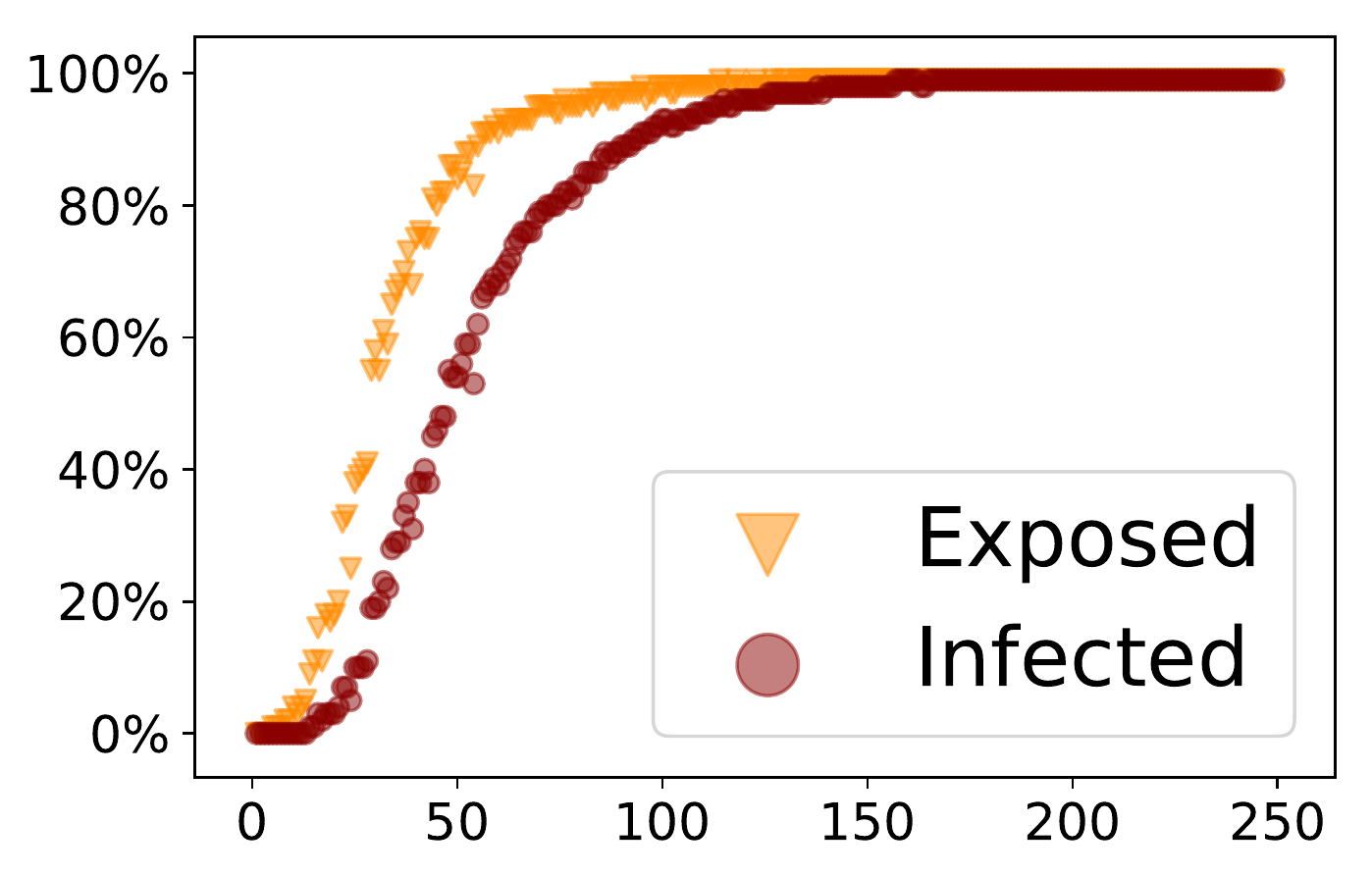}
		\vspace{-0.5cm}
		\caption{India}
		\label{fig:sei_decay_india}
	\end{subfigure}\hspace{-0.15cm}
	\begin{subfigure}[c]{0.28\textwidth}
		\includegraphics[width=\textwidth]{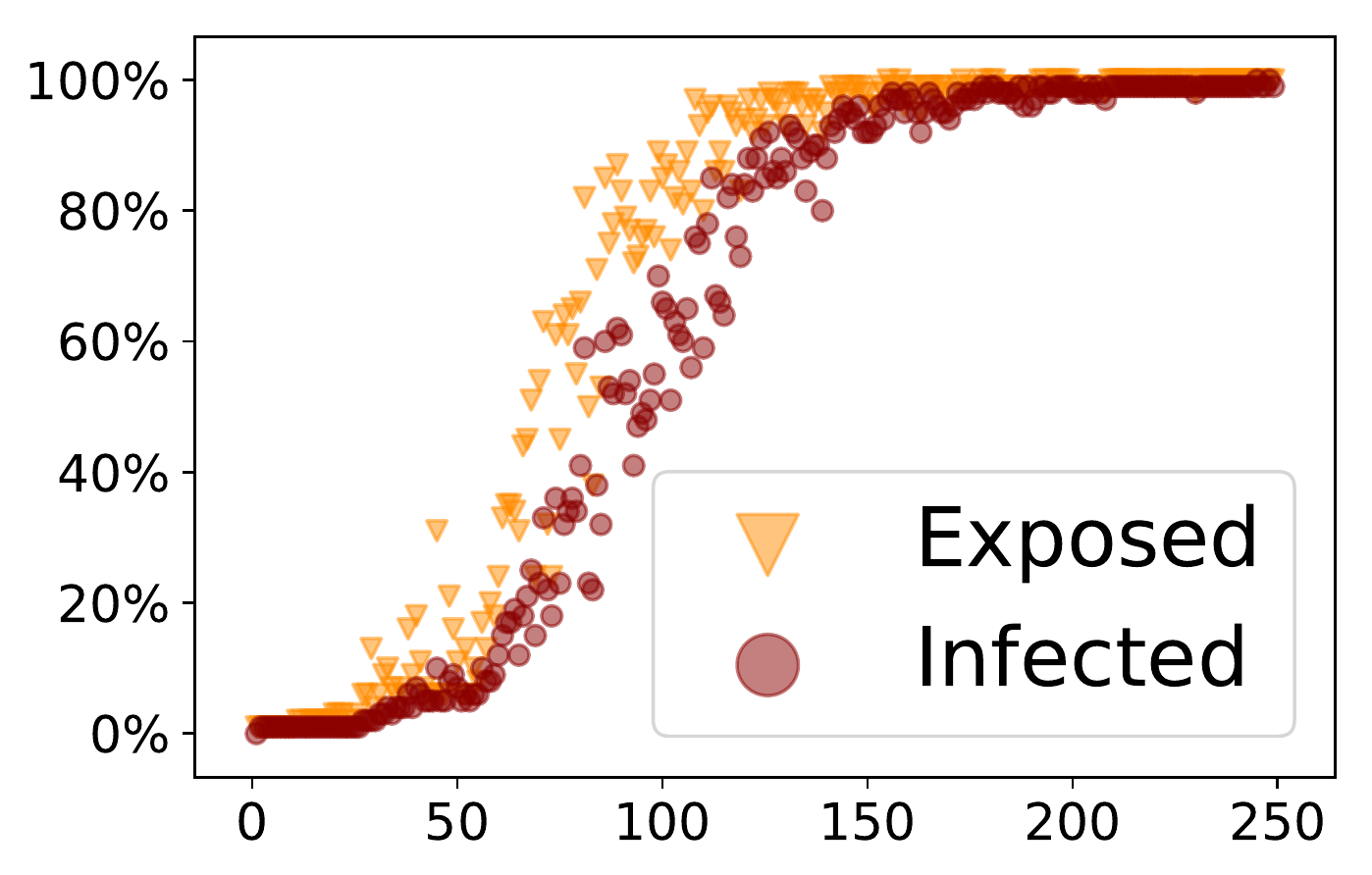}
		\vspace{-0.5cm}
		\caption{Indonesia}
		\label{fig:sei_decay_indonesia}
	\end{subfigure}
	\vspace{-0.2cm}
	\caption{Users infected by time in simulations of the SEI model using max lifetime for infections. $(\alpha) = (\beta) = 0.1$. Forward limit $(\varphi) = 5$.}
	\label{fig:sei_decay}
\vspace{-\baselineskip}
%	\vspace{-0.55cm}
\end{figure}

\textbf{Adding a Max Lifetime to the Infection.}
In reality, users may lose interest in some topics through time, so it is natural for a time limit on the content spread, i.e., content circulates until it loses attention and stagnates. 
We add this time limit to our SEI model, calling this period ``lifetime'', which denotes the maximum duration of an infection in the simulation before it is entirely extinguished. 
Figure \ref{fig:sei_decay} shows the percentage of users infected by increasing the lifetime of the infection. Each data point in the plot indicates a simulation where we fixed the values $\alpha,\beta$ and increased the lifetime an infection could last.
We observe that for all three countries, an infectious content that lasts 100 iterations or more is powerful enough to expose more than half population. When this content persists in the network for at least 150 iterations, it usually infects almost 100\% of the users. 
Note that there is a window of possibility to identify infectious misinformation already spreading (say, around 50 iterations), where a large enough sample of the users were exposed to the content but were not infected and nullify its virality (e.g. disabling forwarding on that piece of content), thus preventing further contagion.

\begin{figure}[t]
	\centering %\hspace{-0.1cm}
	\begin{subfigure}[c]{0.3\textwidth}
		\includegraphics[width=\textwidth]{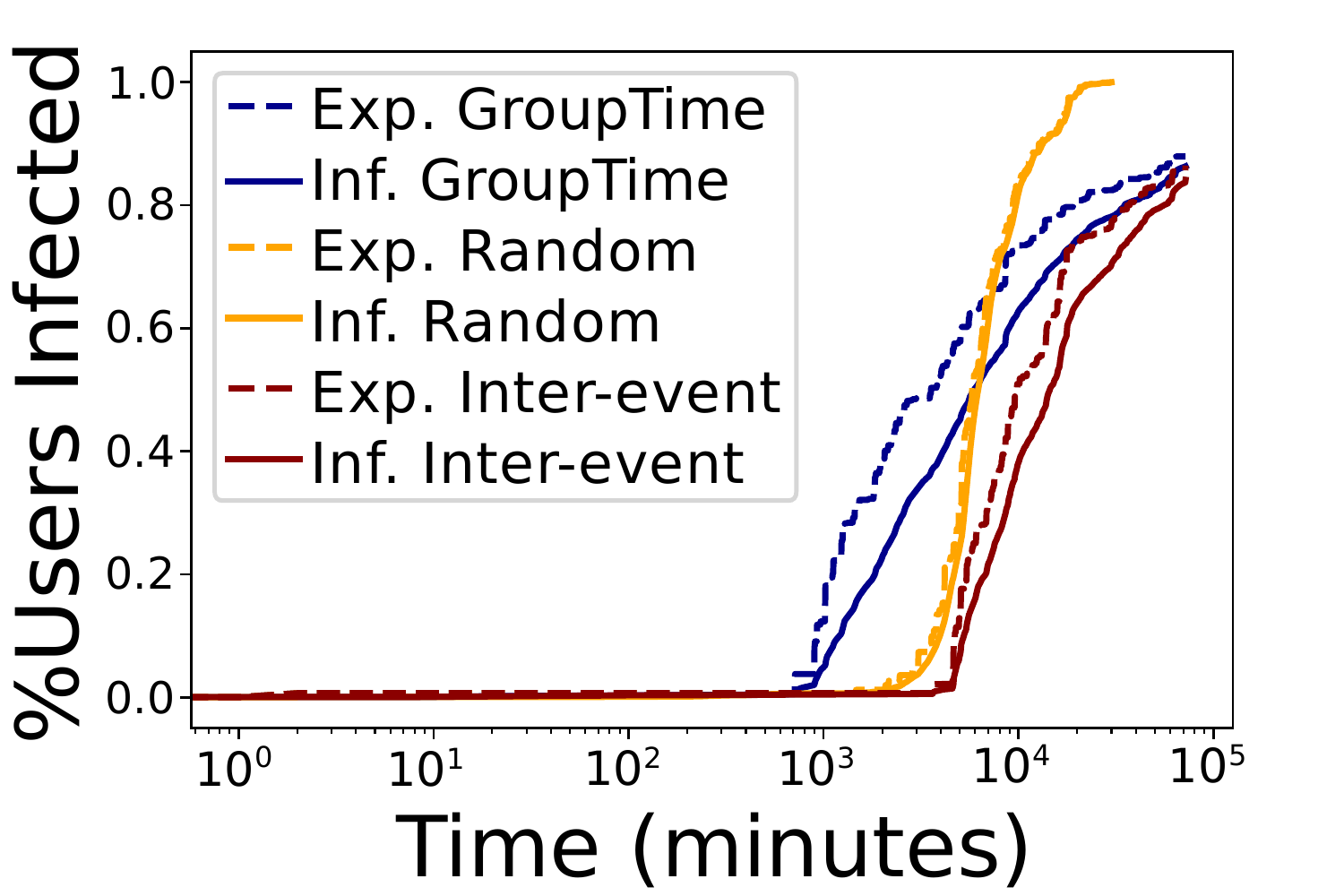} \vspace{-0.55cm}
		\caption{Brazil}
		\label{fig:sei_realtime_brazil}
	\end{subfigure}\hspace{-0.2cm}
	\begin{subfigure}[c]{0.3\textwidth}
		\includegraphics[width=\textwidth]{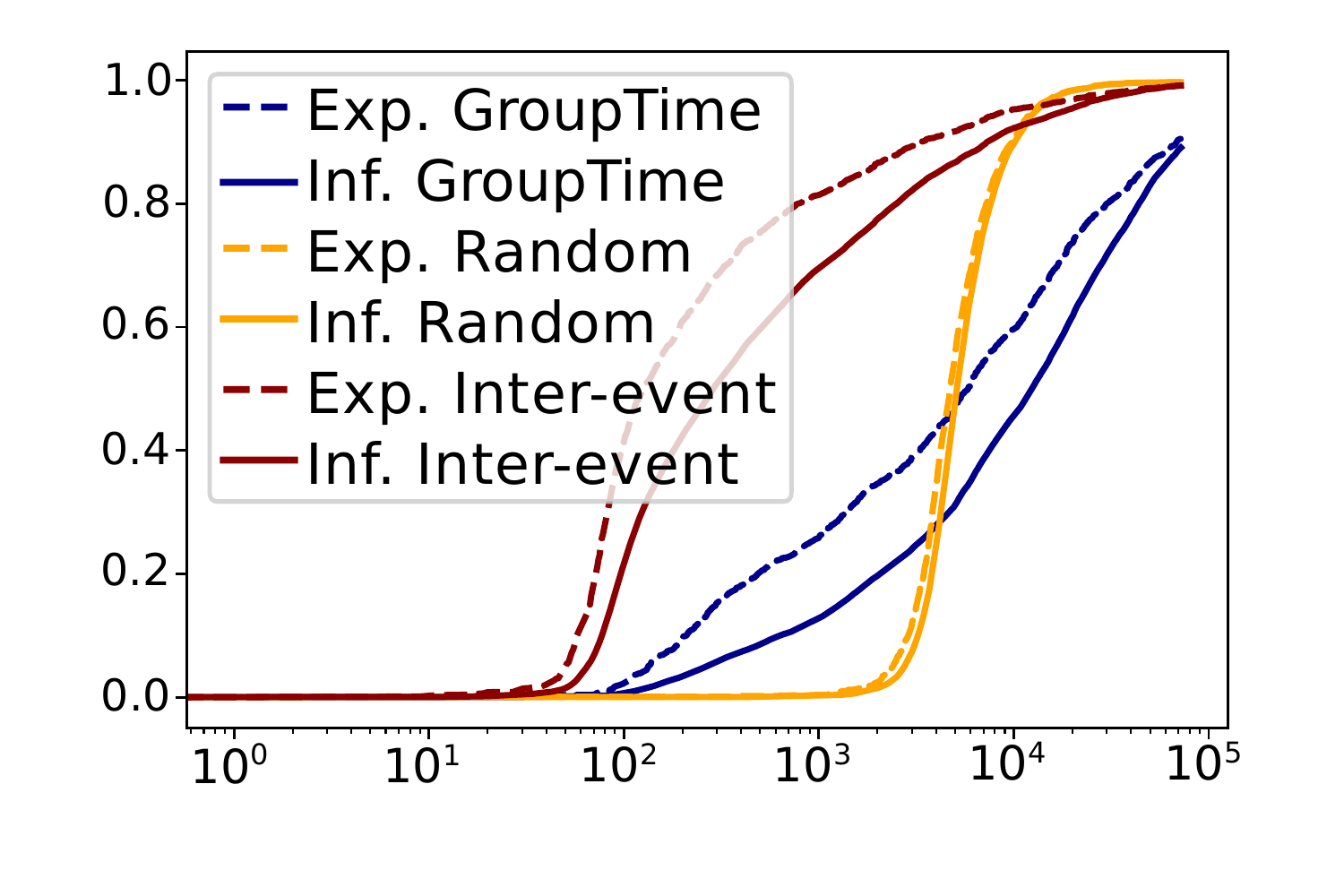} \vspace{-0.55cm}
		\caption{India}
		\label{fig:sei_realtime_india}
	\end{subfigure}\hspace{-0.2cm}
	\begin{subfigure}[c]{0.3\textwidth}
		\includegraphics[width=\textwidth]{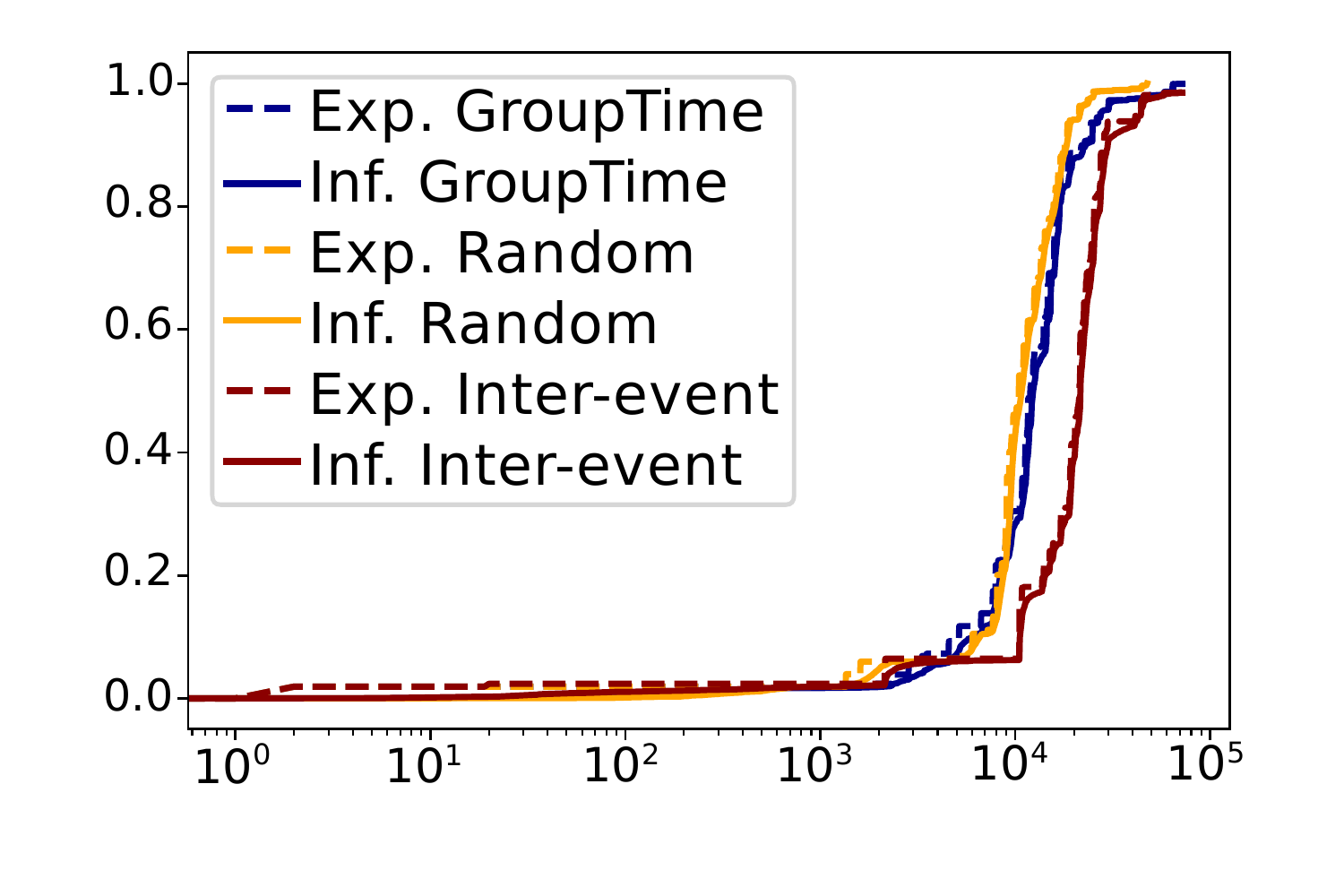} \vspace{-0.55cm}
		\caption{Indonesia}
		\label{fig:sei_realtime_indonesia}
	\end{subfigure}
	\vspace{-0.2cm}
	\caption{Real Time SEI model using ``incubation time'' before spread infection and each iteration equals 1 minute (log). $(\alpha) = (\beta) = 0.1$. Forward limit $(\varphi) = 5$.}
	\label{fig:sei_realtime}
	\vspace{-0.5cm}
\end{figure}

\textbf{Setting Real Time Metrics in SEI Model.} In the previous sections using the SEI model, the spread of information was measured in terms of the \textit{number of iterations}. In this section, we use real data to adapt the SEI model and measure the spread in terms in terms of minutes. 
For this, we add an ``incubation time'' based on the time real data takes to spread over the network. In this version of the model, each iteration represents 1 minute, but when an infected node intends to spread, it has to wait a specific amount of time before doing it. This time is sampled from a distribution of ``waiting times'', which can be: 
(i) \textbf{Random}: a uniform distribution with domain between 1 and 1440 minutes (1 day);
(ii) \textbf{Inter-event Time}: the empirical distribution of inter-event times computed in Figure \ref{fig:burstime2}; 
(iii) \textbf{Group Time}: this strategy is based on the following idea -- it usually takes longer for a message to reach 100 groups than to reach 2 groups. To implement this, in this strategy, we make the incubation time on initial steps smaller than in the subsequent steps. 
During the simulation, we track the number of times the infection has already spread and, for each step, we have a different time distribution according to how long it took for the actual images in WhatsApp to reach those number of groups in our data. 
Figure \ref{fig:sei_realtime} shows experiments considering the three strategies to compute the time to spread. In India, where we have the bursty inter-event times, we see that with the \textit{inter-event time} strategy 60\% of users are exposed to the content in the first 200 minutes of infection. In Brazil, \textit{group time} is faster than \textit{inter-event time} and infected around half of user in the first 2 day (3000 minutes). Finally, in Indonesia all three strategies have very similar behavior, taking over 2 weeks to infect more than 80\% of the users. Nevertheless, a content is still viral when all three strategies are considered, i.e., a misinformation can spread in most of the network before one month of infection.

\vspace{-0.1cm}
\section{Conclusions}
\label{sec:conclusions}
\vspace{-0.1cm}

The closed nature of WhatsApp and the ease of transferring multimedia and sharing information to large-scale groups makes WhatsApp an extremely hard environment for the deployment of countermeasures to combat misinformation. WhatsApp opens a paradoxical use of its platform, allowing at the same time the viral spread of a content and encrypted personal chat. 
Together those two features can be widely abused by misinformation campaigns. 

Our results show that a content can spread quite fast through the network structure of public groups in WhatsApp, reaching later the private groups and individual users. Our empirical observations about the network of WhatsApp public groups in three different countries provides a means of inferring the information velocity in terms of minutes related to real-world scenarios. 
We verified that most of the images (80\%) last no more than 2 days in WhatsApp which, in India, can be already enough to infect half of users in public groups, although there are still 20\% of messages with a time span sufficient to be viral in the three countries using any of our strategies to estimate time of infection.

Using a SEI model we investigate a set of what-if questions about the limits that WhatsApp can impose in the information propagation.  While the limit on the number of users per groups can prohibit the creation of giant hubs to spread information through the network, this limit, however, is not able to prevent a content to reach a large portion of entire platform. More important, our analysis show that low limits imposed on message forwarding and broadcasting (e.g. up to five forwards) offer a delay in the message propagation of up to two orders of magnitude in comparison with the original limit of 256 used in the first version of WhatsApp. 
We note, however, that depending on the virality of the content, those limits are not effective in preventing a message to reach the entire network quickly. 
Misinformation campaigns headed by professional teams with an interest in affecting a political scenario might attempt to create very alarming fake content, that has a high potential to get viral~\cite{resendewww19}. Thus, as a counter-measurement, WhatsApp could implement a quarantine approach to limit infected users to spread misinformation. This could be done by temporarily restricting the virality features of suspect users and content, especially during elections, preventing coordinated campaigns to flood the system with misinformation.

\vspace{-0.2cm}
\bibliographystyle{IEEEtran}
{\small
\bibliography{ref}
}
\end{document}